\documentclass[UTF8, twocolumn, times]{aastex62}
\hypersetup{linkcolor=magenta,citecolor=cyan,filecolor=green,urlcolor=blue}

\usepackage{graphicx}	
\usepackage{amsmath}	
\usepackage{amssymb}	

\usepackage{epsfig}
\usepackage{multirow}
\usepackage{float}
\usepackage{footmisc}

\usepackage{enumitem}
\setlist[itemize]{leftmargin=*}

\def\mnras{MNRAS}
\def\apj{ApJ}

\def\apjl{ApJL}
\def\apjs{ApJS}
\def\aap{A\& A}

\def\nat{Nature}
\def\sgra{Sgr A$^\star$}
\def\xmm{{\it XMM-Newton}}
\def\cxo{{\it Chandra}}

\def\axj{AX J1745.6-2901}
\def\swf{SWIFT J1658.2-4242}

\accepted{for Publication in ApJ}

\shorttitle{Probing ISM through an Asymmetric Dust Scattering Halo}
\shortauthors{Jin et al.}

\begin{document}

\title{Exploring the Interstellar Medium Using an Asymmetric X-ray Dust Scattering Halo}

\correspondingauthor{Chichuan	 Jin}
\email{ccjin@nao.cas.cn}

\author[0000-0002-2006-1615]{Chichuan Jin}
\affiliation{National Astronomical Observatories, Chinese Academy of Sciences, A20 Datun Road, Beijing 100101, China}

\author{Gabriele Ponti}
\affiliation{INAF, Osservatorio Astronomico di Brera Merate, via E. Bianchi 46, I-23807 Merate, Italy}
\affiliation{Max-Planck-Institut f\"{u}r Extraterrestrische Physik, Giessenbachstrasse, D-85748 Garching, Germany}

\author{Guangxing Li}
\affiliation{South-Western Institute for Astronomy Research, Yunnan University, Kunming 650500, China}

\author{David Bogensberger}
\affiliation{Max-Planck-Institut f\"{u}r Extraterrestrische Physik, Giessenbachstrasse, D-85748 Garching, Germany}



\begin{abstract}

\swf\ is an X-ray transient discovered recently in the Galactic plane, with severe X-ray absorption corresponding to an equivalent hydrogen column density of $N_{\rm H,abs}\sim2\times10^{23}$ cm$^{-2}$. Using new \cxo\ and \xmm\ data, we discover a strong X-ray dust scattering halo around it. The halo profile can be well fitted by the scattering from at least three separated dust layers. During the persistent emission phase of \swf, the best-fit dust scattering $N_{\rm H,sca}$ based on the COMP-AC-S dust grain model is consistent with $N_{\rm H,abs}$. The best-fit halo models show that 85-90 percent of the intervening gas and dust along the line of sight of \swf\ are located in the foreground ISM in the Galactic disk. The dust scattering halo also shows significant azimuthal asymmetry, which appears consistent with the inhomogeneous distribution of foreground molecular clouds. By matching the different dust layers to the distribution of molecular clouds along the line of sight, we estimate the source distance to be $\sim$10 kpc, which is also consistent with the results given by several other independent methods of distance estimation. The dust scattering opacity and the existence of a halo can introduce a significant spectral bias, the level of which depends on the shape of the instrumental point spread function and the source extraction region. We create the {\sc xspec} {\tt dscor} model to correct for this spectral bias for different X-ray instruments. Our study reenforces the importance of considering the spectral effects of dust scattering in other absorbed X-ray sources.

\end{abstract}



\section{Introduction}
\subsection{X-ray Transient: \swf}
\label{sec-swf}
\swf\ is an X-ray transient whose first detection was reported on 2018 February 16 in  {\it Swift}/BAT (\citealt{Barthelmy.2018}; \citealt{DAvanzo.2018}; \citealt{Lien.2018}). It was also detected by {\it INTEGRAL} on 2018 February 14 (\citealt{Grebenev.2018}; \citealt{Grinberg.2018}). \citet{Lien.2018} refined the analysis of the {\it Swift}/XRT data and reported the source coordinates being Ra=$16^{\rm h}58^{\rm m}12^{\rm s}.64$, Dec=$-42\degr41'54.4''$, with the 90\% uncertainty being 1.6". This position corresponds to the Galactic coordinates $l=343.25153,~b=+0.05387$, indicating that the source should be in the Galactic plane, at an angular separation of 16.693 degree from \sgra. They also found that the {\it Swift}/XRT spectrum of \swf\ was well fitted by an absorbed power law model, with the photon index being $1.7\pm1.5$ and a large neutral hydrogen column density for the X-ray absorption: $N_{\rm H,abs}~=~(1.9\pm0.5)\times10^{23}$ cm$^{-2}$. These spectral properties were also confirmed by \citet{Beri.2018} using an {\it AstroSat} observation of 20 ks exposure time on 2018 February 20. They found that the photon index was $1.76\pm0.06$ and $N_{\rm H,abs}=(1.6\pm0.2)~\times10^{23}$ cm$^{-2}$.

\citet{RussellTD.2018} reported radio observation of the source on 2018 February 17 using Australia Telescope Compact Array (ATCA). They found that the source position in the radio band was Ra=$16^{\rm h}58^{\rm m}12^{\rm s}.700$ ($\pm0.004$), Dec=$-42\degr41'56.09''$ ($\pm0.25$). This radio position is consistent with the X-ray position. \citet{RussellDM.2018} reported optical observations between 2018 February 22 and 25 in the SDSS $\sc i$  and $\sc r$ bands, using an one-meter telescope in the Las Cumbres Observatory. An optical source was detected at the radio position, with the $i$ band AB magnitude being 19.07 $\pm$ 0.06 $mag$, which was the only source within 8 arcsec of the radio source. But no significant optical variability was found in this optical source during the X-ray outburst of the transient, so they concluded that this optical source was likely to be a foreground star, rather than being the true optical counterpart.

Most recently, \citet{Xu.2018} reported results from a full spectral-timing analysis of 33.3 ks {\it NuSTAR} exposure on 2018 February 16. They found that the broadband X-ray continuum could be well fitted by a cut-off power law plus a Compton reflection. In their best-fit model 4, they reported $N_{\rm H,abs}=(1.80\pm0.08)~\times10^{23}$ cm$^{-2}$, photon index was $1.63\pm0.02$ and the coronal temperature was $22\pm1$ keV. The black hole spin was found to be larger than 0.96. The binary system was found to have a large inclination angle of $64^{+2}_{-3}$ degree, which was consistent with the detection of dips in the {\it NuSTAR} X-ray light curve. The $N_{\rm H,abs}$ during the dips increased to $1.00^{+0.08}_{-0.07}~\times10^{24}$ cm$^{-2}$. Type-C Quasi-Periodic Oscillation (QPO) signals were also detected in the X-ray power spectrum. All the above results suggest that \swf\ should be a typical black hole X-ray binary in the hard state, and is viewed at a high inclination angle.

\begin{table*}
   \caption{List of \cxo\ and \xmm\ DDT observations on \swf.}
    \label{tab-obs}
\begin{tabular}{cccccccccc}
\hline
\hline
Satellite & OBSID & Obs-Date & Instrument & Exp & $\theta_{\rm offaxis}$ & $E_{\rm 2-4keV}$ & $E_{\rm 4-6keV}$ & $E_{\rm 6-10keV}$ & $F_{\rm 2-10keV}$ \\
 & & & & (ks) & (arcmin) & (keV) & (keV) & (keV) & (erg cm$^{-2}$ s$^{-1}$) \\
\hline
\cxo\ & 21083 & 2018-04-28 & ACIS-S & 28.9 & 0.391 & 3.3 & 4.9 & 6.9 & $0.62\times10^{-9}$ \\
\hline
\xmm\ & 0811213401 &  2018-02-27 & MOS1 & 25.0 & 1.702 & 3.1 & 4.8 & 6.9 & $2.62\times10^{-9}$ \\
\xmm\ & 0805200201a &  2018-03-04 & MOS2 & 7.9 & 1.704 & 3.2 & 4.8 & 7.0 & $3.72\times10^{-9}$ \\
\xmm\ & 0805200201b &  2018-03-04 & MOS2 & 8.7  & 1.704 & 3.2 & 4.8 & 7.0 & $2.50\times10^{-9}$ \\
\xmm\ & 0805200301 &  2018-03-11 & MOS2 & 44.5  & 1.690 & 3.2 & 4.8 & 6.9 & $1.48\times10^{-9}$ \\
\xmm\ & 0805200401 &  2018-03-15 & MOS2 & 36.5  & 1.733 & 3.2 & 4.8 & 6.9 & $1.72\times10^{-9}$ \\
\xmm\ & 0805201301 &  2018-03-28 & MOS2 & 31.5 & 1.700 & 3.2 & 4.8 & 6.9 & $1.61\times10^{-9}$ \\
\hline
\end{tabular}
\\
\\
{\bf Notes.} {\it Exp} is the exposure time after all the data filtering steps (see Section~\ref{sec-reduction}). $\theta_{\rm offaxis}$ is the off-axis angle relative to the optical axis of the instrument. $E_{\rm 2-4keV}$, $E_{\rm 4-6keV}$ and $E_{\rm 6-10keV}$ are the spectral weighted effective energy in the 2-4, 4-6 and 6-10 keV band, respectively. $F_{\rm 2-10keV}$ is the observed (absorbed) mean flux of the source in the 2-10 keV band. 0805200201a and 0805200201b are the high and low flux time intervals in the observation OBSID: 0805200201 (see Section~\ref{sec-xmm}).
\end{table*}

\subsection{X-ray Dust Scattering Halo}
Dust grains in the interstellar medium (ISM) can scatter X-ray photons with small scattering angles. This phenomenon was predicted half a century ago (\citealt{Overbeck.1965}; \citealt{Trumper.1973}). The scattering can be calculated by the Mie scattering theory. Provided that there is an X-ray point-like source and a dust layer in front of it, dust scattering will create a small halo around the source (\citealt{Mathis.1991}). There are also calculations about the time delay of the scattered photons in the halo due to different light paths from those photons coming directly from the line-of-sight (LOS) (\citealt{Xu.1986}). Since the first observational discovery of a dust scattering halo around the source GX339-4 by the {\it Einstein} satellite (\citealt{Rolf.1983}), such effect has been repeatedly observed in several tens of X-ray sources (e.g. \citealt{Predehl.1995}; \citealt{Xiang.2005}; \citealt{Valencic.2015}). The timing effects of dust scattering have also been observed in several X-ray sources, with short bursts of X-ray emission, therefore showing ring-like structures around them (e.g. \citealt{Beardmore.2016}; \citealt{Heinz.2016}; \citealt{Vasilopoulos.2016}), as well as sources undergoing eclipses showing typical eclipse light curves (\citealt{Jin.2018}).

It was also noticed long time ago that because of the existence of a halo, the radial profile of a point-like X-ray source should be different from the standard point spread function (PSF) of an instrument (\citealt{Predehl.1992}). However, dust scattering opacity is often much smaller than the X-ray absorbing opacity, thus it was often ignored in previous X-ray spectral studies. For example, dust scattering opacity is not considered in the widely used {\sc xspec} (v12.10.1, \citealt{Arnaud.1996}) X-ray absorbing models such as {\tt phabs} and {\tt tbabs} (\citealt{Wilms.2000}). \citet{Corrales.2016} reported that ignoring the effects of dust scattering opacity could lead to an over-estimate of $N_{\rm H, abs}$ by a baseline level of 25\%. \citet{Smith.2016} pointed out that a proper treatment of the dust scattering should consider both the type of the dust grains and the source extraction region. \citet{Jin.2017} showed that for the X-ray binary \axj\ in the Galactic Centre (GC) with $N_{\rm H,abs}\sim3\times10^{23}$ cm$^{-2}$ (\citealt{Ponti.2018}), dust scattering created severe spectral biases in terms of both spectral shape and flux. These biases depended on the shape of the source extraction region and the instrumental PSF. It is also clear that dust scattering significantly affect the spectra of most (likely all) sources at the GC (Jin et al. in preparation).

Since \swf\ lies in the Galactic plane and its $N_{\rm H,abs}$ is on the order of $10^{23}$ cm$^{-2}$, X-ray dust scattering in its LOS can be strong and complex. Therefore, it is both important and interesting to study the X-ray dust scattering halo and check if it can lead to severe spectral biases. In this paper, we use the latest \cxo\ and \xmm\ observations to study the properties of the X-ray dust scattering halo around \swf, and explore the significance of the spectral biases. The paper is organized as follows. We first describe the observation and data reduction in Section 2, and then present detailed study of the dust scattering halo in Section 3. The analysis of the halo asymmetry is presented in Section 4. Then in Section 5 we discuss the foreground ISM distribution, the source distance, the azimuthal asymmetry of the halo shape, and the spectral biases. The final section summarizes the main conclusions of our study.

\section{Observation and Data Reduction}
\label{sec-reduction}
This work is based on the data from a 29.9 ks \cxo\ Director's Discretionary Time (DDT) observation conducted on 28th April 2018, using the spectroscopic array of the Advanced CCD Imaging Spectrometer (ACIS-S) instrument with the High-Energy Transmission Grating (HETG). It also uses data from 5 \xmm\ DDT observations conducted between 27th February and 28th March 2018 (Table~\ref{tab-obs}), each with one Metal-Oxide-Silicon (MOS) camera in the Full Frame mode (Medium filter) and the other two European Photon Imaging Cameras (EPIC) in the Timing mode (Bogensberger et al. in preparation). The data reduction procedures are described below.

\begin{figure}
\centering
\includegraphics[bb=0 0 612 640, clip=1, scale=0.3]{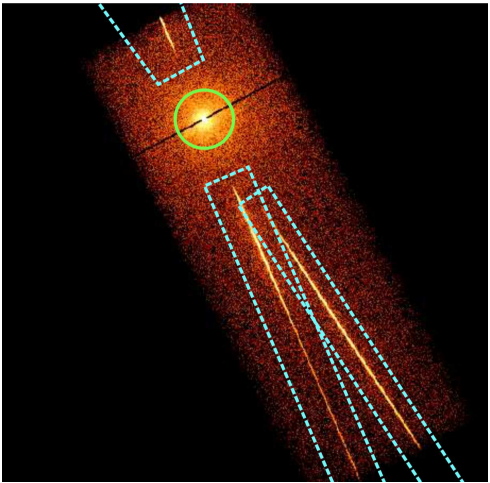} 
\caption{The count image of \swf\ in the ACIS-S2 chip. Regions shown by the cyan dash lines were masked out to minimize the contamination from HETG arms. For comparison, a green circle with a radius of 30 arcsec is put at the source position. The dark line across the source is due to the subtraction of the readout streak.}
\label{fig-mask}
\end{figure}

\begin{figure}
\centering
\includegraphics[clip=, bb=0 0 700 665, scale=0.34]{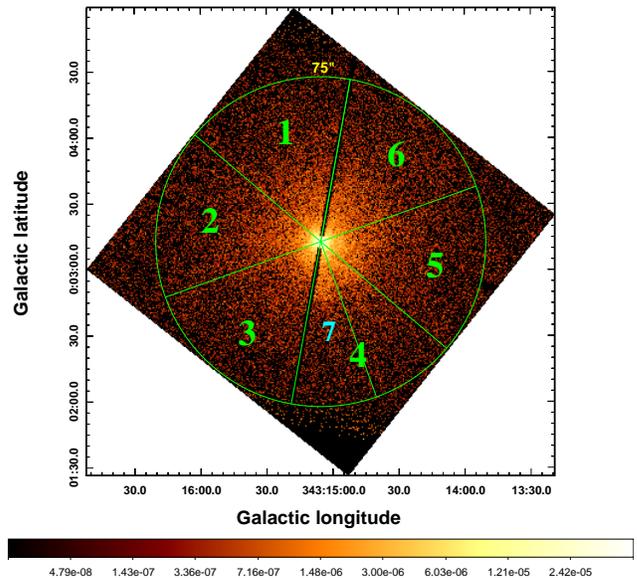} 
\caption{The flux image observed in \cxo\ ACIS-S in 0.5-7 keV band. The readout streak was removed. The bright pixels around \swf\ indicate the X-ray dust scattering halo. The region was divided into several sub-regions where the radial profiles were extracted separately for detailed analysis (see Section~\ref{sec-assy}).}
\label{fig-region}
\end{figure}

\begin{figure*}
\centering
\includegraphics[clip=1, bb=220 0 504 792, scale=0.62,angle=90]{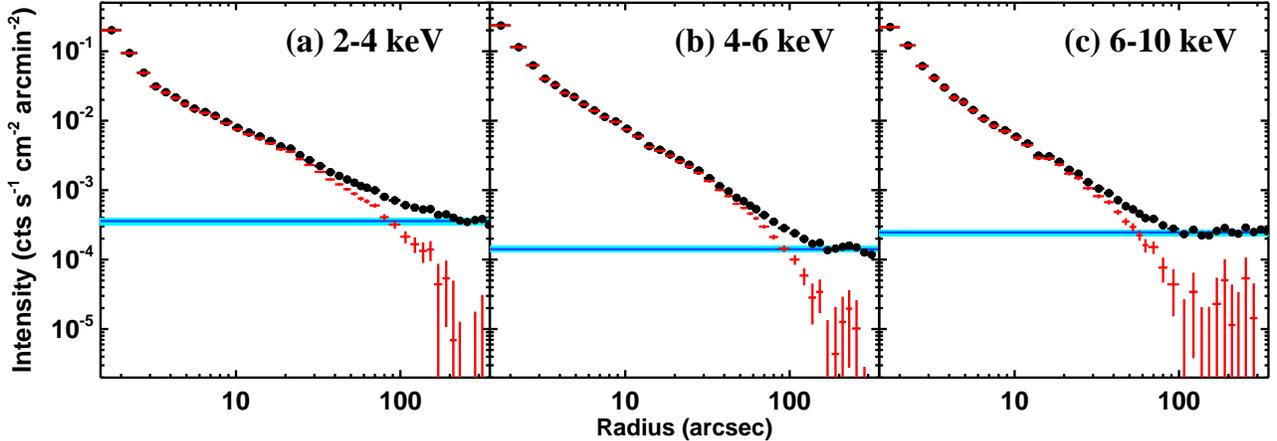} 
\caption{The radial profile of \swf\ before the background subtraction (black data points) and after the background subtraction (red data points). The background flux (blue line, the $\pm1\sigma$ range was shown in cyan) was determined from the average flux outside the radius when the radial profile turned flat, which was about 220, 150 and 110 arcsec for 2-4, 4-6 and 6-10 keV, respectively.}
\label{fig-bkgsub}
\end{figure*}

\subsection{\cxo}
\label{sec-cxo}
We used the Chandra Interactive Analysis of Observations ({\sc CIAO}) software (v4.10, \citealt{Fruscione.2006}) and the latest Calibration Data Base (CALDB, v4.7.9) to perform the data reduction. Firstly the {\tt chandra\_repro} script was used to reprocess the data. Then we followed the thread on the Chandra website to remove all the background flaring periods using the {\tt deflare} script\footnote{\url{http://cxc.cfa.harvard.edu/ciao/threads/flare/}}. This resulted in a new event file with totally 28.9 ks clean exposure time.

Then the {\tt acisreadcorr} script was used to remove artificial features due to the readout streak produced by out-of-time events. To ensure a clean removal of these features, the streak width was chosen to be 5 pixels, instead of the default value of 2 pixels, as shown in Figure~\ref{fig-mask}. The remaining out-of-time events from the halo were negligible compared to the primary halo emission even at large radii, so that no extra steps were taken to remove them.
Since the source was on the back-illuminated ACIS-S3 CCD chip, we excluded the data on the two nearby front-illuminated CCD chips (i.e. ACIS-S2 and S4) because of their different background count rates. Then the {\tt fluximage} script was used to create flux image and exposure maps. The {\tt wavdetect} script was used to perform point source detection within the field-of-view. The off-axis angle of \swf\ relative to the optical axis was found to be 0.391 arcmin during the observation. We extracted the light curve of \swf\ and found that during this \cxo\ observation the source did not show any flip-flop behaviour as seen in \xmm\ observations (see Section~\ref{sec-xmm}), thus it was not necessary to exclude any data for the variability issue. By using the {\tt pileup\_map} script, we found that the central circular region around the source with 2.5 arcsec radius had more than 1\% pile-up level. 

Moreover, bright features along the HETG arms arise from the dispersion of source photons could seriously contaminate the primary halo intensity. We visually inspected the counts image of the chip, and masked out large areas around these features to minimize their contamination, as shown in Figure~\ref{fig-mask}. Then the {\tt funtools} in {\sc ds9} (v7.5, \citealt{Joye.2003}) was used to extract values within a set of annuli around the source in order to obtain the radial profile. This azimuthal averaging also helps to smear out remaining contaminations.

Similar to \citet{Jin.2017,Jin.2018}, we studied the halo profile in three energy bands: 2-4, 4-6 and 6-10 keV. This band division could reveal the energy dependence of the halo, meanwhile keeping enough signal-to-noise in every band. Then we calculated the spectral-weighted effective energy in every band (Table~\ref{tab-obs}), which was used to calculate the halo model in that band. The central 20 arcsec of the PSF was obtained from simulations using the \cxo\ online {\sc ChaRT} PSF simulator. The PSF wing outside 20 arcsec was calculated using the analytical formulas in \citet{Gaetz.2010}. These PSF profiles were added to the radial profile modelling and to be convolved with the dust scattering halo.

\begin{figure*}
\centering
\includegraphics[clip=1, bb=220 0 520 792, scale=0.62, angle=90]{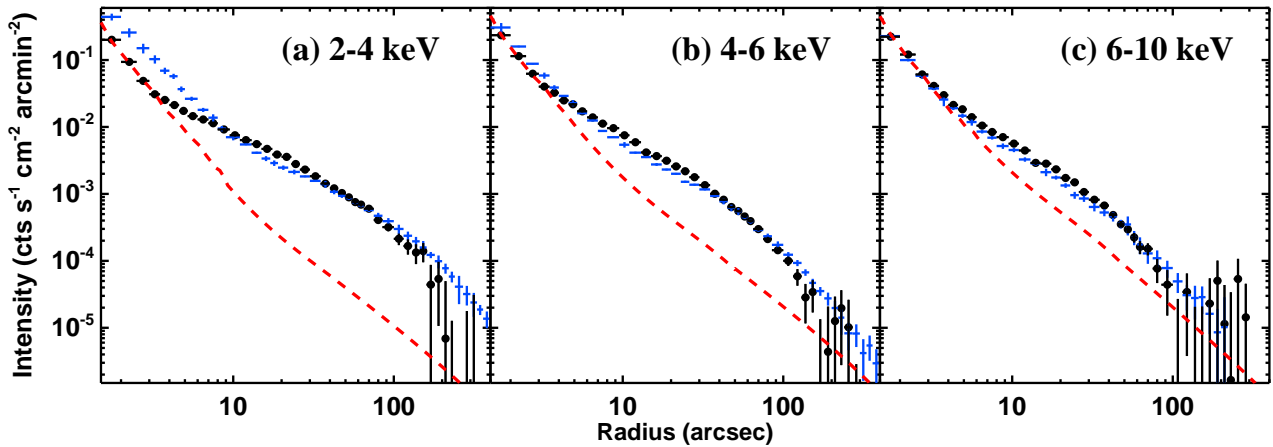} 
\caption{Background subtracted radial profiles of \swf\ in the three energy bands observed by \cxo\ ACIS-S, as shown by the black points. The red dash line is the PSF normalized to the flux of \swf\ within 3 arcsec. For comparison, the blue data points show the composite radial profile of X-ray sources within the central 2 square degree of the GC, renormalized to the flux of \swf\ within the radial range of 60-80 arcsec (Jin et al. in preparation).}
\label{fig-halocompare}
\end{figure*}

\subsection{\xmm}
\label{sec-xmm}
We used the \xmm\ (\citealt{Jansen.2001}) Science Analysis Software (SAS, v17.0.0, \citealt{Gabriel.2004}), especially the Extended Source Analysis Software (ESAS; \citealt{Snowden.2014}), and the latest calibration files to perform the data reduction. Since \swf\ was bright enough to cause severe pile-up, the EPIC-pn camera and one MOS camera were put in the Timing mode, while the other MOS camera was operating in the Full-Frame mode in order to detect the dust scattering halo. In this work about the dust scattering halo, we only made use of the MOS data in the imaging mode. The data reduction was conducted as follows. Firstly, the {\tt odfingest}, {\tt emchain} and {\tt mos-filter} scripts were run in sequence to reprocess the data and obtain a new event file. Then both the source and background light curves were extracted to check the high background periods and source variability. The source showed intriguing flux jumps and dips during the \xmm\ observations. These phenomena are not related to the topic of this work (see Bogensberger et al. in preparation for detailed analysis), but they may cause some variability in the halo shape (\citealt{Jin.2018}). Therefore, we excluded all the dips, and carefully chose time intervals to avoid flux jumps in order to ensure the stability of the halo. Since the source was very bright ($F_{\rm 2-10~keV}\sim10^{-9}$ erg cm$^{-2}$ s$^{-1}$ observed during all the observations in Table~\ref{tab-obs}), signal-to-noise was not a problem for the \xmm\ data even for a short exposure time of only a few kilo-seconds.

In OBSID: 0805200201, the source showed a flip-flop behaviour appearing as high-flux and low-flux platforms, the flux difference between them was $\gtrsim$ 20\% in 2-10 keV. We decided to extract halo profiles from the two platforms separately. The first time interval contained 7.9 ks clean exposure in the high-flux platform, and the second interval contained 8.7 ks clean exposure in the low-flux platform. In OBSID: 0805201301, the first 11 ks and the interval within 18-21 ks showed significant flux drops of $\sim$ 10\% in 2-10 keV, while the final 13 ks showed enhanced variability of $\sim$ 10\% in the same band, thus these periods were excluded, only the remaining 25.0 ks clean exposure was used. In OBSID: 0811213401, the first 26 ks and last 3 ks contained dramatic flux jumps of more than a factor of 1.5 in 2-10 keV, so these periods were excluded, and the total clean exposure left for the halo extraction was 31.5 ks (see Table~\ref{tab-obs}, light curves and detailed spectral-timing analysis can be found in Bogensberger et al. in preparation). There was no flip-flop behaviour in the rest two \xmm\ observations, so all the data were used.

For the clean data within these time intervals, we used {\tt mos-spectra} and {\tt mos\_back}  scripts to create counts images, flux images and exposure maps in the 2-4, 4-6 and 6-10 keV bands. Then the {\tt cheese} script was run to detect serendipitous point sources in the field-of-view to be masked out. Finally, the {\tt funtools} in {\sc ds9} was used to obtain the radial profiles of \swf\ in the three energy bands. Within each energy band, we calculated the spectral-weighted effective energy (Table~\ref{tab-obs}), which was used as input for the halo modelling and PSF production. As in \citet{Jin.2017,Jin.2018}, we used the \citet{Ghizzardi.2002} analytical PSF profiles for the halo model convolution and radial profile analysis.

\begin{table*}
\begin{minipage}{175mm}
  \centering
\caption{Best-fit parameters for different scenarios of foreground dust distribution (see Section~\ref{sec-halofit}).}
   \label{tab-halofit}
\begin{tabular}{lcccccccccc}
\hline
\hline
Layer & Parameter & Scenario-1 & Scenario-2 & Scenario-3 & Scenario-4 & Scenario-5 & Unit \\ 
\hline
Layer-1 & x$_{\rm low,1}$ & 0.000$^{+0.001}_{l}$ & 0.990$^{u}_{-0.002}$ & 0.990$^{u}_{-0.003}$ & 0.990$^{u}_{-0.002}$ & 0.990$^{u}_{-0.003}$ & \\
& x$_{\rm high,1}$ & 0.888$^{+0.005}_{-0.005}$ & 1.000$^{u}_{-0.019}$ & 0.996$^{+0.002}_{-0.002}$ & 1.000$^{+0.020}_{-0.020}$ & 1.000$^{u}_{-0.010}$ & \\
& $f_{\rm nH,1}$ & 100.0$^{u}$ & 14.1$^{+5.2}_{-5.2}$ & $13.3^{+1.3}_{-1.3}$ & $13.2^{+1.3}_{-1.3}$ & $13.3^{+1.3}_{-1.3}$ & \% \\
\hline
Layer-2 & x$_{\rm low,2}$ & -- & 0.000$^{+0.001}_{l}$ & 0.666$^{+0.062}_{-0.062}$ & 0.867$^{+0.037}_{-0.037}$ & 0.7-fixed & \\
& x$_{\rm high,2}$ & -- & 0.911$^{+0.005}_{-0.005}$ & 0.906$^{+0.019}_{-0.019}$ & 0.867$^{+0.037}_{-0.037}$ & 0.9-fixed & \\
& $f_{\rm nH,2}$ & -- & 85.9$^{+2.8}_{-2.8}$ & 25.8$^{+3.0}_{-3.0}$ & 11.3$^{+1.0}_{-1.0}$ & 23.7$^{+3.8}_{-3.8}$ & \% \\
\hline
Layer-3 & x$_{\rm low,3}$ & -- & -- & 0.000$^{+0.390}_{l}$ & 0.697$^{+0.095}_{-0.095}$ &0.5-fixed & \\
& x$_{\rm high,3}$ & -- & --& 0.401$^{+0.052}_{-0.052}$ & 0.697$^{+0.095}_{-0.095}$ & 0.7-fixed & \\
& $f_{\rm nH,3}$ & -- & --& 60.9$^{+9.2}_{-9.2}$ & 17.7$^{+4.6}_{-4.6}$ & 0.0$^{+21.6}_{l}$ & \% \\
\hline
Layer-4 & x$_{\rm low,4}$ & -- & --& -- & 0.007$^{+0.064}_{-0.007}$ & 0.3-fixed & \\
& x$_{\rm high,4}$ & -- & --& -- & 0.335$^{+0.226}_{-0.226}$ & 0.5-fixed & \\
& $f_{\rm nH,4}$ & -- & --& -- & 57.8$^{+6.2}_{-6.2}$ & 17.7$^{+4.4}_{-4.4}$ & \% \\
\hline
Layer-5 & x$_{\rm low,5}$ & -- & -- & -- & -- & 0.1-fixed & \\
& x$_{\rm high,5}$ & -- & -- & -- & -- & 0.3-fixed & \\
& $f_{\rm nH,5}$ & -- & -- & -- & -- & 26.6$^{+8.9}_{-8.9}$ & \% \\
\hline
Layer-6 & x$_{\rm low,6}$ & -- & -- & -- & -- & 0.0-fixed & \\
& x$_{\rm high,6}$ & -- & -- & -- & -- & 0.1-fixed & \\
& $f_{\rm nH,6}$ & -- & -- & -- & -- &18.7$^{+3.0}_{-3.0}$ & \% \\
\hline
& $N_{\rm H,tot}$ & 1.03$^{+0.01}_{-0.01}$ & 1.55$^{+0.09}_{-0.09}$ & 1.66$^{+0.16}_{-0.16}$ & 1.66$^{+0.13}_{-0.13}$ & 1.65$^{+0.40}_{-0.40}$ & $10^{23}$ cm$^{-2}$ \\
\hline
& $\chi^2_{\rm \nu}$ & 348.3/96 & 214.3/93 & 108.0/90 & 107.3/87 & 108.1/91 & \\
\hline
\end{tabular}
\end{minipage}
\\
\\
{\bf Notes.} Error bars indicate 1$\sigma$ parabolic errors. Scenario-1, 2, 3, 4, 5 comprise 1, 2, 3, 4, 6 dust layers, respectively. All dust layers adopt the COMP-AC-S dust grain population. $u$ and $l$ signify that the parameter pegged at its upper and lower limit during the minimal $\chi^2$ run. `fixed' indicates that the parameter value was fixed during the halo profile fitting.
\end{table*}

\section{Analysis of the X-ray Dust Scattering Halo}
\subsection{Shape of the Azimuthal-averaged Halo}
\label{sec-haloshape}
Both \cxo\ ACIS and \xmm\ MOS observed a strong halo around \swf, but the MOS data were heavily contaminated by the pile-up effect, so we present the results from MOS in Appendix~\ref{app-sec-mos}, and only used the ACIS data of much higher angular resolution for more detailed halo profile analysis. Firstly we created the flux image of \swf\ in ACIS-S in 0.5-7 keV band (the default `broad' energy band in the {\tt fluximage} script), as shown in Figure~\ref{fig-region}. The source does not appear exactly like a point source, instead it is surrounded by bright pixels with the intensity decreasing towards large radii. This is a clear indication for the existence of a strong halo. Azimuthal variation of the halo intensity is also apparently visible in the figure, which we discuss in more detail in Section~\ref{sec-assy}.

Then we extracted the azimuthal-averaged radial profiles from 2-4, 4-6 and 6-10 keV bands, separately. Since there was no archival \cxo\ observation on the same position of \swf\ before its outburst, it was not possible to directly measure the X-ray background, including both X-ray diffuse emission and detector background underneath the halo of \swf, as was done for another transient \axj\ in the GC (\citealt{Jin.2017}). Therefore, we assumed a flat background underneath the halo profile. The background flux was derived from the average intensity outside the radius where the radial profile becomes flat, which is about 220, 150 and 120 arcsec in 2-4, 4-6 and 6-10 keV band, respectively (Figure~\ref{fig-bkgsub}). The subtraction of this background has little effect on the radial profile within $\sim$30 arcsec, where the source is more than one order of magnitude brighter than the background. But due to the limitation of background subtraction and source brightness, the halo profile is significantly detectable only at $\lesssim$ 200 arcsec. Thus it is not possible to check the existence of an extended halo wing as seen in \axj\ (\citealt{Jin.2017,Jin.2018}).

Figure~\ref{fig-halocompare} shows the background subtracted radial profiles of \swf\ within the three energy bands. The PSF profile, renormalized to the source flux within 3 arcsec, is over-plotted for comparison. Note that a dust component local to the source will create a scattering halo with a similar profile as the PSF even within 3 arcsec, which is difficult to disentangle using halo profile study alone (\citealt{Jin.2018}). Excess flux is clearly observed outside 3.5 arcsec, and the excess decreases with the increasing energy. This is a strong evidence for the presence of an X-ray dust scattering halo. For comparison, we also over-plot the composite radial profile observed in \cxo\ ACIS for X-ray sources within 2 square degrees of \sgra\ in the GC (Jin et al. in preparation), where the $N_{\rm H, abs}$ is also on the order of $10^{23}$ cm$^{-2}$. By renormalizing them to the profile of \swf\ within the radial range of 60-80 arcsec, we see that the halo of \swf\ appears very different from the GC composite halo. This implies that the foreground dust distribution towards \swf\ should be quite different from the GC direction. Thus although the dust distribution may be relatively uniform within the central 2 degrees around \sgra\ (Jin et al. in preparation), it is significantly different on scales of 10-20 degrees.

\begin{figure*}[ht]
\begin{tabular}{c}
\includegraphics[trim=0in 1.9in 0in 0in, clip=1, bb=0 0 792 465, scale=0.6]{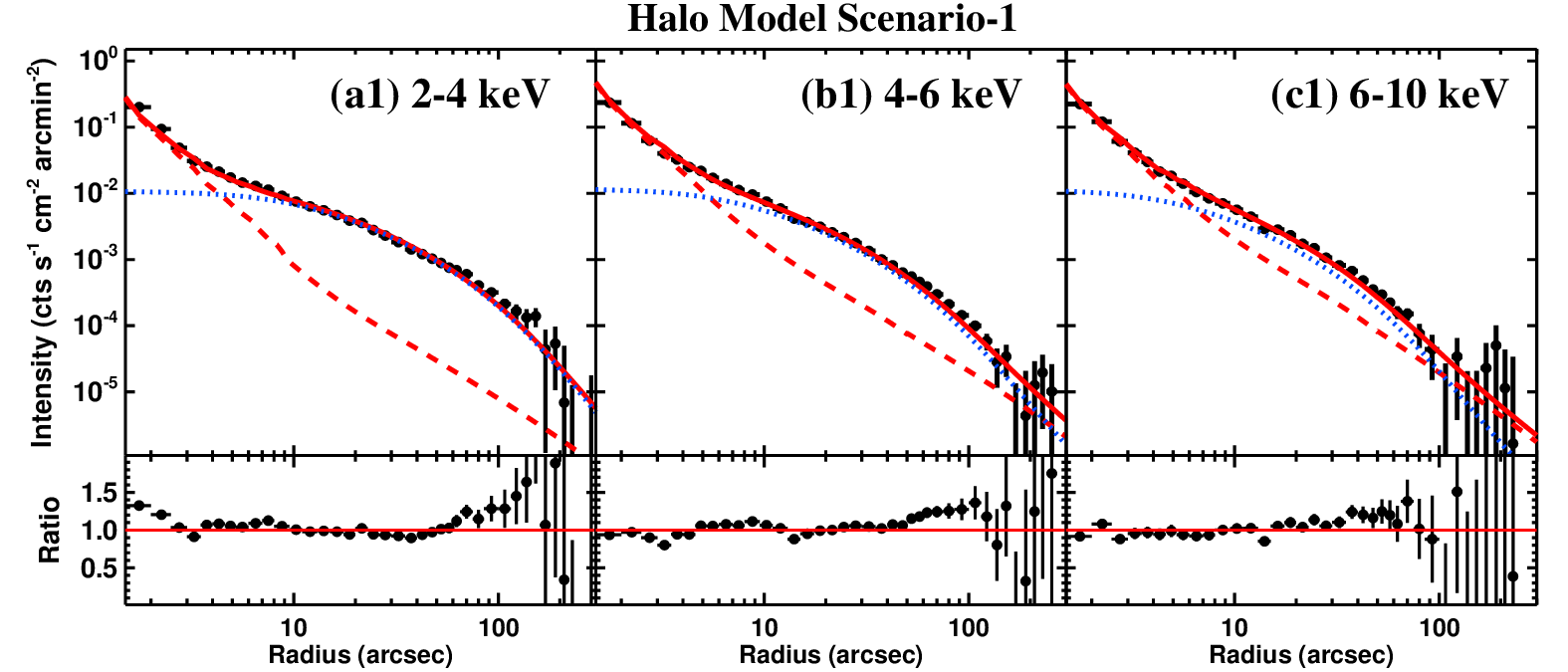} \\
\includegraphics[trim=0in 1.9in 0in 0in, clip=1, bb=0 0 792 465, scale=0.6]{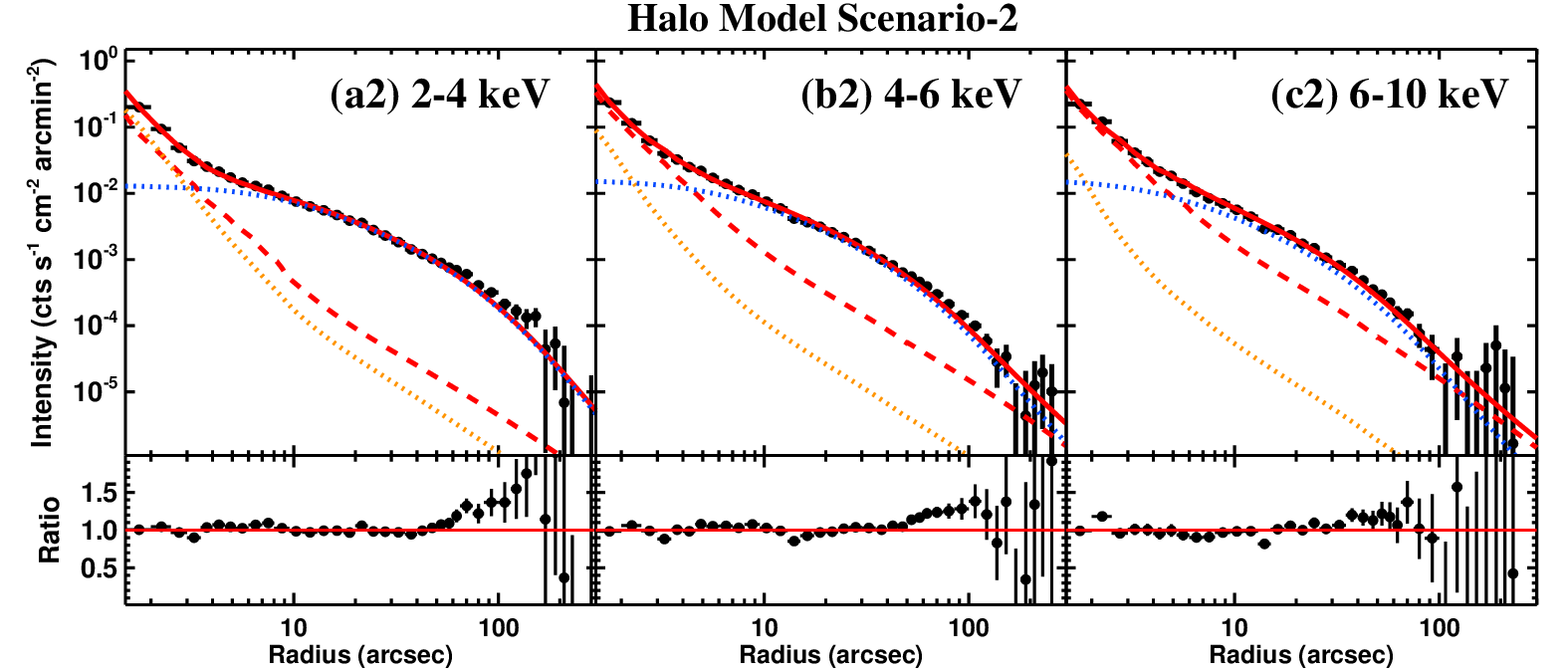} \\
\includegraphics[trim=0in 1.7in 0in 0in, clip=1, bb=0 0 792 465, scale=0.6]{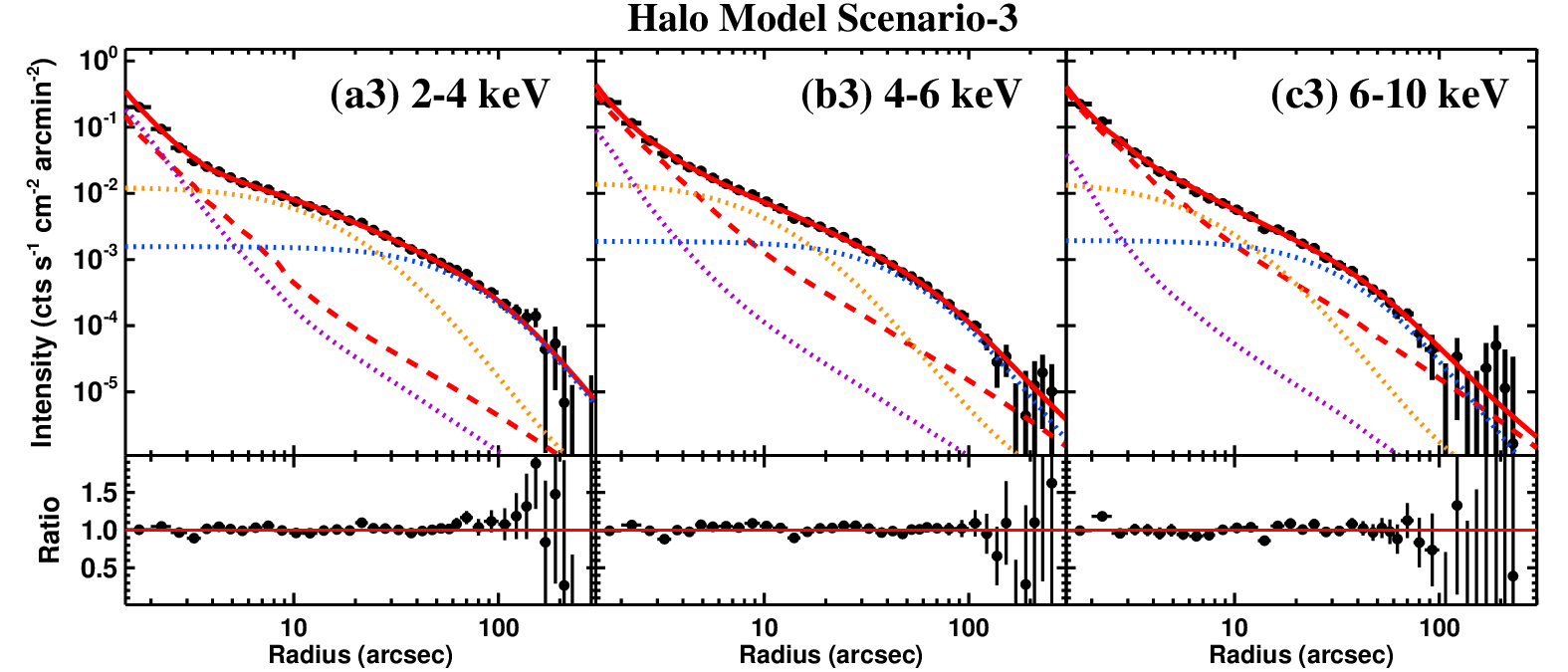} \\
\end{tabular}
\caption{Best-fit results for the halo profile modelling of \swf, using three different scenarios of foreground dust distribution. In each panel, the black points show the background subtracted radial profile. The red solid line is the total model, the red dash line is the PSF. All the dotted lines represent different halo components, with more extended components originating from closer dust layers to Earth. The best-fit parameters are listed in Table~\ref{tab-halofit}.}
\label{fig-halofit}
\end{figure*}
\addtocounter{figure}{-1}
\begin{figure*}
\begin{tabular}{c}
\includegraphics[trim=0in 1.9in 0in 0in, clip=1, bb=0 0 792 465, scale=0.6]{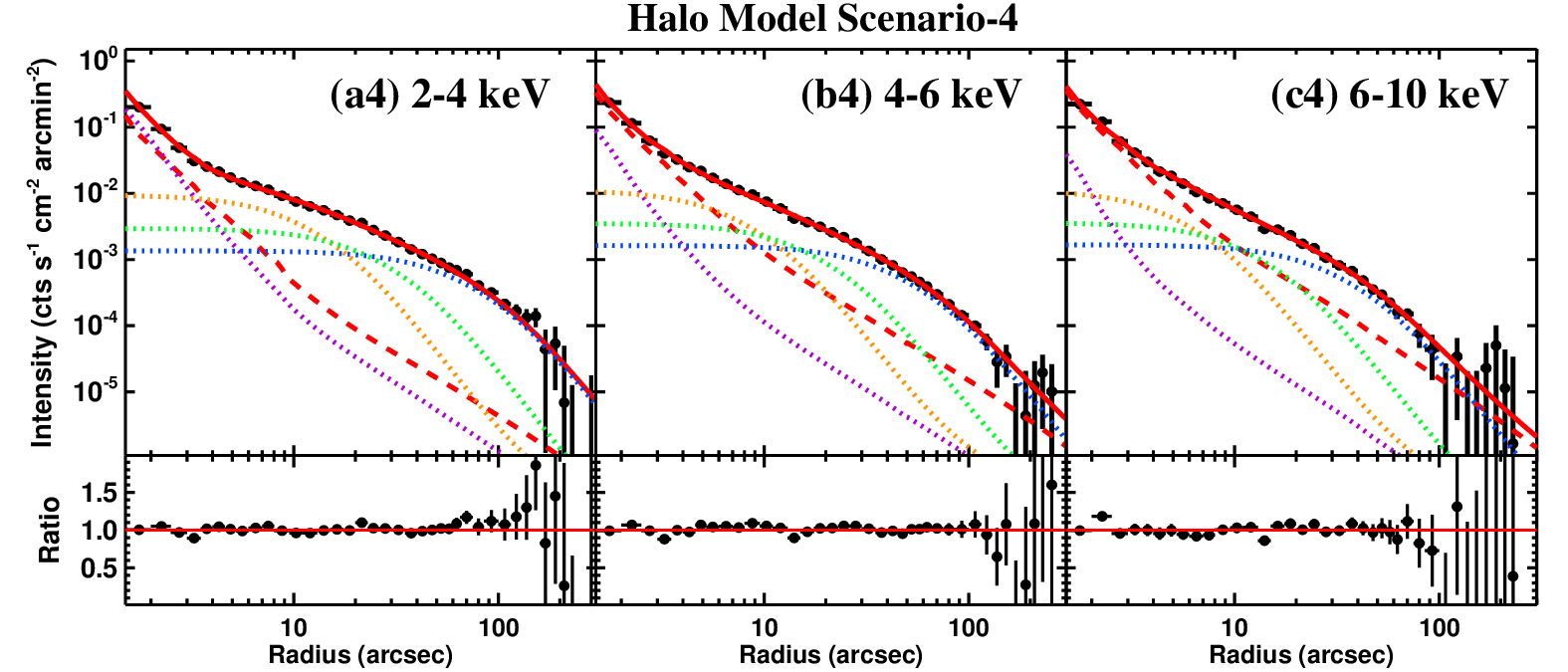} \\
\includegraphics[trim=0in 1.7in 0in 0in, clip=1, bb=0 0 792 465, scale=0.6]{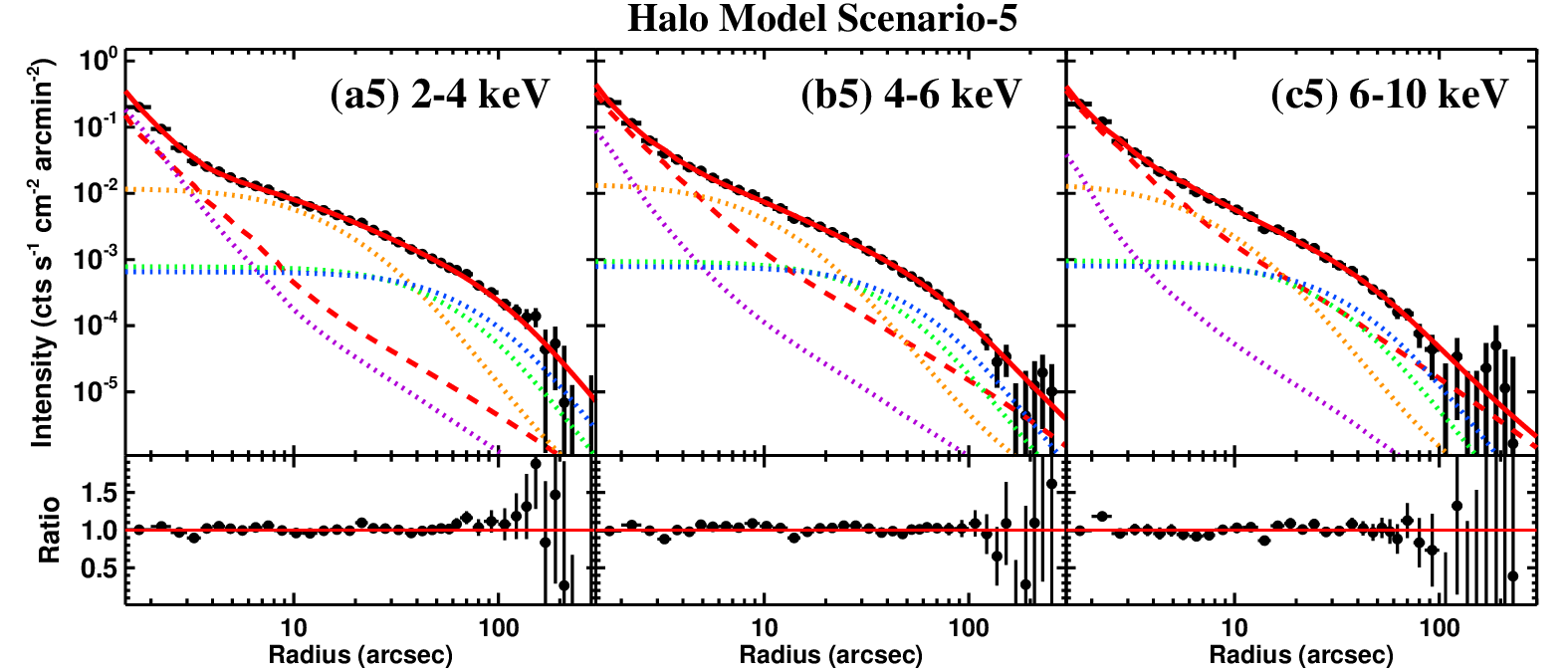} \\
\end{tabular}
\caption{{\it continuted}}
\label{fig-halofit2}
\end{figure*}

\subsubsection{Modelling the Azimuthal-averaged Halo Profile}
\label{sec-halofit}
In order to model the halo of \swf, we applied the method in \citet{Jin.2017} which used several dust scattering components from multiple geometrically-thick foreground dust layers to fit the halo profile. We used the dimensionless fractional distance ($x$) to calculate the halo shape, thus the halo modelling is not affected by the source distance which might be uncertain (see Section~\ref{sec-discussion1}). $x$ is defined between 0 and 1, with the observer being at 0 and source being at 1. Since there is no knowledge about the number of major dust layers along the LOS, we tried five different model scenarios for the LOS dust distribution. The type of dust grain is also a major source of uncertainty. We chose three typical dust grain populations reported in \citet{Zubko.2004}. The first one is the COMP-AC-S dust grain population, which has been recommended for the GC direction (\citealt{Fritz.2011}). Since \swf\ is only 16 degree from \sgra, it is possible that COMP-AC-S is also a reasonably good approximation for it. The other two are BARE-GR-B and COMP-NC-B. \citet{Jin.2017} showed that among all the 19 dust grain populations tried in their work, these two populations could create the most extended and most compact halos, separately. Thus the best-fit parameters based on these two grain populations can reflect the range of parameter dispersion due to the choice of different grain populations. The {\tt Minuit} algorithm in the Python {\sc iminuit} (v1.3.3, \citealt{James.1975}) interface was used to perform the minimum $\chi^2$ fitting. The best-fit parameters are given in Tables~\ref{tab-halofit}, \ref{app-tab-halofit1} and \ref{app-tab-halofit2}.

We began the halo profile modelling with the COMP-AC-S grain population. In Scenario-1, which contains only one geometrically thick dust layer, the lower and upper boundaries and the $N_{\rm H,sca}$ (the equivalent hydrogen column density for the dust scattering) of the layer are all free parameters. The best-fit model shows that the layer distributes within $0 \le x \le 0.888$, with the total $N_{\rm H,sca}$ being $(1.03\pm0.01)~\times10^{23}$ cm$^{-2}$. But the $\chi^2$ is only 348.3 for 96 degrees-of-freedom (DOFs). The bad $\chi^2$ is confirmed by the significant residuals shown in the Panels a1, b1 and c1 in Figure~\ref{fig-halofit}.

\begin{figure*}
\includegraphics[clip=1, bb=35 30 1062 530, scale=0.5]{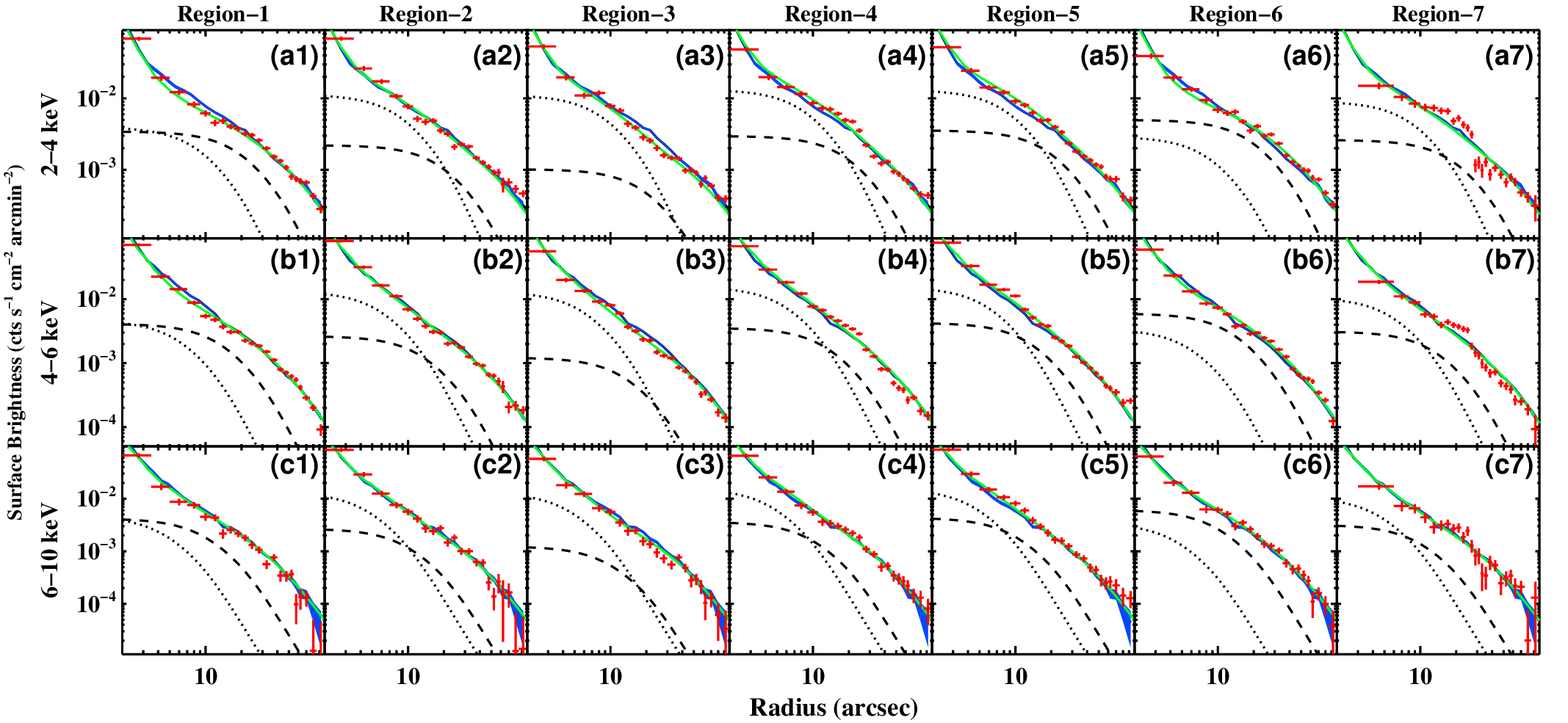}
\caption{The radial profiles of \swf\ observed in \cxo\ ACIS-S in different azimuthal regions as shown in Figure~\ref{fig-region}. In each panel, the red data points show the source radial profile in the sub-region. The blue curve indicates the azimuthal-averaged radial profile including the 1$\sigma$ error as shown in Figure~\ref{fig-halocompare}. The best-fit halo model (green solid line) was derived by changing only the $N_{\rm H,sca}$ of the halo components from layer-2 (black dotted line) and layer-3 (black dash line). The halo components from layer-1, layer-4 and PSF are not plotted as they are the same as in the Panels a4, b4 and c4 in Figure~\ref{fig-halofit}.}
\label{fig-assy-radp}
\end{figure*}

\begin{table*}
  \begin{minipage}{180mm}
  \centering
   \caption{Best-fit $N_{\rm H,sca}$ of each dust layer in Scenario-4 for different azimuthal subregions.}
    \label{tab-obs-assy}
\begin{tabular}{lcccccccc}
\hline
\hline
 & $N_{\rm H,sca}$ & $f_{\rm region-1}$ & $f_{\rm region-2}$ & $f_{\rm region-3}$ & $f_{\rm region-4}$ & $f_{\rm region-5}$ & $f_{\rm region-6}$ & $f_{\rm region-7}$ \\
& ($10^{22}$ cm$^{-2}$) & \\
\hline
Layer-1 & $2.2$ & 1.00-fixed & 1.00-fixed& 1.00-fixed& 1.00-fixed& 1.00-fixed& 1.00-fixed& 1.00-fixed\\
Layer-2 & $1.9$ & $0.41\pm0.08$ & $1.18\pm0.09$ & $1.17\pm0.08$ & $1.39\pm0.09$ & $1.37\pm0.09$ & $0.30\pm0.09$ & $0.95\pm0.11$ \\
Layer-3 & $3.0$ & $1.16\pm0.07$ & $0.74\pm0.08$ & $0.34\pm0.07$ & $1.00\pm0.08$ & $1.19\pm0.08$ & $1.68\pm0.08$ & $0.88\pm0.10$ \\
Layer-4 & $9.6$ & 1.00-fixed & 1.00-fixed& 1.00-fixed& 1.00-fixed& 1.00-fixed& 1.00-fixed& 1.00-fixed\\
\hline
All Layers & $16.7$ & $0.96\pm0.02$ & $0.97\pm0.02$ & $0.90\pm0.02$ & $1.04\pm0.02$ & $1.08\pm0.02$ & $1.04\pm0.02$ & $0.97\pm0.02$ \\
\hline
     \end{tabular}
\end{minipage}
\\
{\bf Notes.} $f_{\rm region-n}$ signifies the best-fit $N_{\rm H,sca}$ of each layer in region-$\it n$ relative to the best-fit $N_{\rm H,sca}$ of each layer in the azimuthal-averaged halo profile. The $N_{\rm H,sca}$ of layer-1 and layer-4 was fixed during the fitting (see Section~\ref{sec-assy}). The last row shows the ratio between the total $N_{\rm H,sca}$ in each region and the value of the azimuthal-averaged halo.
\end{table*}

Then we tried Scenario-2, which contains two dust layers. Layer-1 is closer to the binary system and layer-2 is closer to Earth. The best-fit model gives $\chi^2_{\rm \nu}=214.3/93$, indicating a significant improvement compared to Scenario-1. Layer-1 is found to distribute within $0.99 \le x \le 1$ and contains ($14.1\pm5.2$)\% LOS dust. The remaining dust is contained in layer-2, which is located within $0 \le x \le 0.911$. The total $N_{\rm H,sca}$ is found to be $(1.55\pm0.09)~\times10^{23}$ cm$^{-2}$. According to the Panels a2, b2 and c2 in Figure~\ref{fig-halofit}, the main improvement is in the central 5 arcsec region, where the halo contribution is more significant in Scenario-2, but there are still severe residuals outside 40 arcsec radius.

Then a third layer was added, namely Scenario-3. In this scenario, the best-fit $\chi^2$ is 105.4 for 90 DOF, indicating a further significant improvement. The best-fit model shows that (13.3$\pm1.3$)\% LOS dust is in layer-1 at $0.990 \le x \le 0.996$, ($25.8\pm3.0$)\% LOS dust is in layer-2 at $0.666 \le x \le 0.906$, and the remaining ($60.9\pm9.2$)\% LOS dust is in layer-3 at $0 \le x \le 0.401$. The total $N_{\rm H,sca}$ is $(1.66\pm0.16)~\times10^{23}$ cm$^{-2}$, which is consistent with the result from Scenario-2. According to the Panels a3, b3 and c3 in Figure~\ref{fig-halofit}, the main model improvement is in the radial range outside 40 arcsec, with the new $\chi^2$ being 108.0 for 90 DOFs.

Scenario-4 contains 4 dust layers. The best-fit halo decomposition is shown in the Panels a4, b4 and c4 in Figure~\ref{fig-halofit}. However, the $\chi^2$ only improves by 0.7 for 3 additional free parameters, thus adding more layers cannot bring further improvement in the halo fitting. The best-fit total $N_{\rm H,sca}$ is also consistent with the result of Scenario-3.

As a further test, we tried Scenario-5 with 6 layers, as shown by the Panels a5, b5 and c5 in Figure~\ref{fig-halofit}. Layer-1 is still local to the source, and another 5 foreground layers distributing within the fractional distance ranges of 0.0-0.1, 0.1-0.3, 0.3-0.5, 0.5-0.7, 0.7-0.9, respectively. The boundaries of these 5 layers were fixed during the fitting, only their $N_{\rm H, sca}$ values were allowed to vary. In the best-fit model of this scenario, ($13.3\pm1.3$)\% LOS dust is contained in layer-1 at $0.99 \le x \le 1$. For the rest 5 layers, ($23.7\pm3.8$)\% LOS dust is located within $0.7 \le x \le 0.9$, 0\% within $0.5 \le x \le 0.7$, ($17.7\pm4.4$)\% within $0.3 \le x \le 0.5$, ($26.6\pm8.9$)\% within $0.1 \le x \le 0.3$, and ($18.7\pm3.0$)\% within $0.0 \le x \le 0.1$. The best-fit $\chi^2$ is 108.1 for 91 DOFs, and the total $N_{\rm H,sca}$ is $(1.65\pm0.40)~\times10^{23}$ cm$^{-2}$, both are consistent with the results in Scenario-3 and 4.

All the error bars come from the 1$\sigma$ parabolic error returned by the {\sc iminuit} code. However, we emphasize that this statistical uncertainty, even derived from the more robust Monte Carlo simulation as done in \citep{Jin.2018}, cannot reflect the true uncertainty of the halo profile fitting. This is because most of the uncertainty should arise from the lack of knowledge about the properties of dust grains, such as their grain type, metallicity, spatial and size distribution. All these physical properties can lead to a significant change of the best-fit values. To assess this uncertainty, we also tried BARE-GR-B and COMP-NC-B grain populations for all the model scenarios. We found that despite the change of best-fit parameters the resultant relative distribution of LOS dust did not change significantly, and Scenario-4 always gave the best $\chi^2$ (see Appendix~\ref{app-sec-halofit}). This is also consistent with \citep{Jin.2017} who tried 19 different dust grain populations and found that the conclusion for the general LOS dust distribution did not change.

Consequently, we can conclude that the observed average halo profile requires at least 3 dust layers to fit,  and adding more dust layers cannot bring further improvement. In all the scenarios with more than 2 layers, the one local to \swf\ contains (10-15)\% LOS dust regardless of the number of layers or the type of dust grains. The majority of LOS dust is in the foreground layers significantly far away from the source and located along the Galactic disk. However, it must be emphasized that this relative dust distribution is still based on the assumption that the dust grain population does not change along the LOS.

Moreover, it is necessary to emphasize that the halo model used in this work does not include multiple scattering. \citet{Mathis.1991} shows that multiple scattering can dominate over single scattering when the scattering optical depth $\tau_{sca}$ is larger than 1.3. The major effect of multiple-scattering is causing the halo profile to be more extended (\citealt{Mathis.1991}; \citealt{Xiang.2007}; \citealt{Jin.2017}). Since we know $\tau_{sca} \propto E^{-2}$ where $E$ is the photon energy (\citealt{Predehl.1995}), it is clear that multiple scattering is much more important at lower energies. Adopting $N_{\rm H,sca}=16.6\times10^{23}$ cm$^{-2}$ from the best-fit halo model in Table~\ref{tab-halofit}, we can calculate $\tau_{sca}$ to be 0.90, 0.41 and 0.21 in the 2-4, 4-6 and 6-10 keV band respectively, thus single scattering should dominate in these bands. This is further supported by the fact that our single-scattering based halo models can already reproduce the observed halo profile and its energy dependence reasonably well. Therefore, we can conclude that multiple-scattering should not affect our results.

\begin{figure*}
\begin{tabular}{c}
\includegraphics[clip=1, bb=0 120 760 450, scale=0.6]{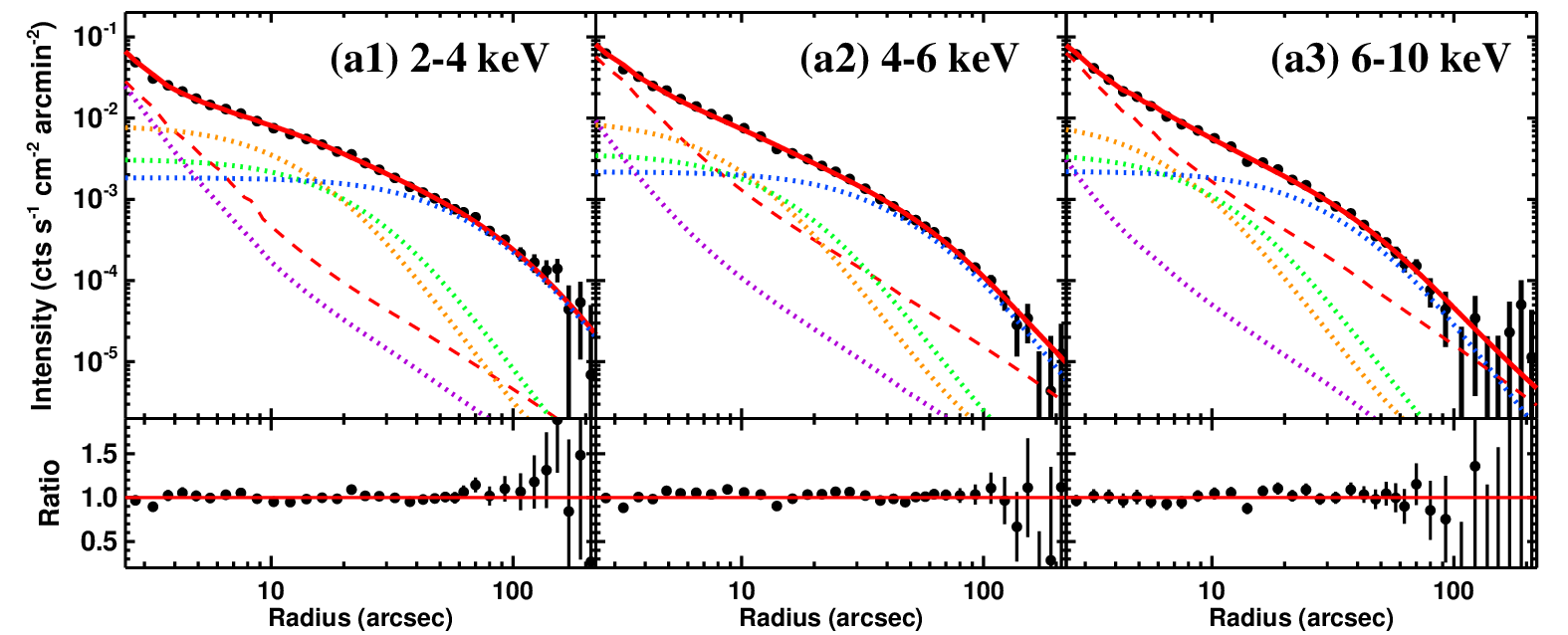}\\
\includegraphics[clip=1, bb=0 120 760 430, scale=0.6]{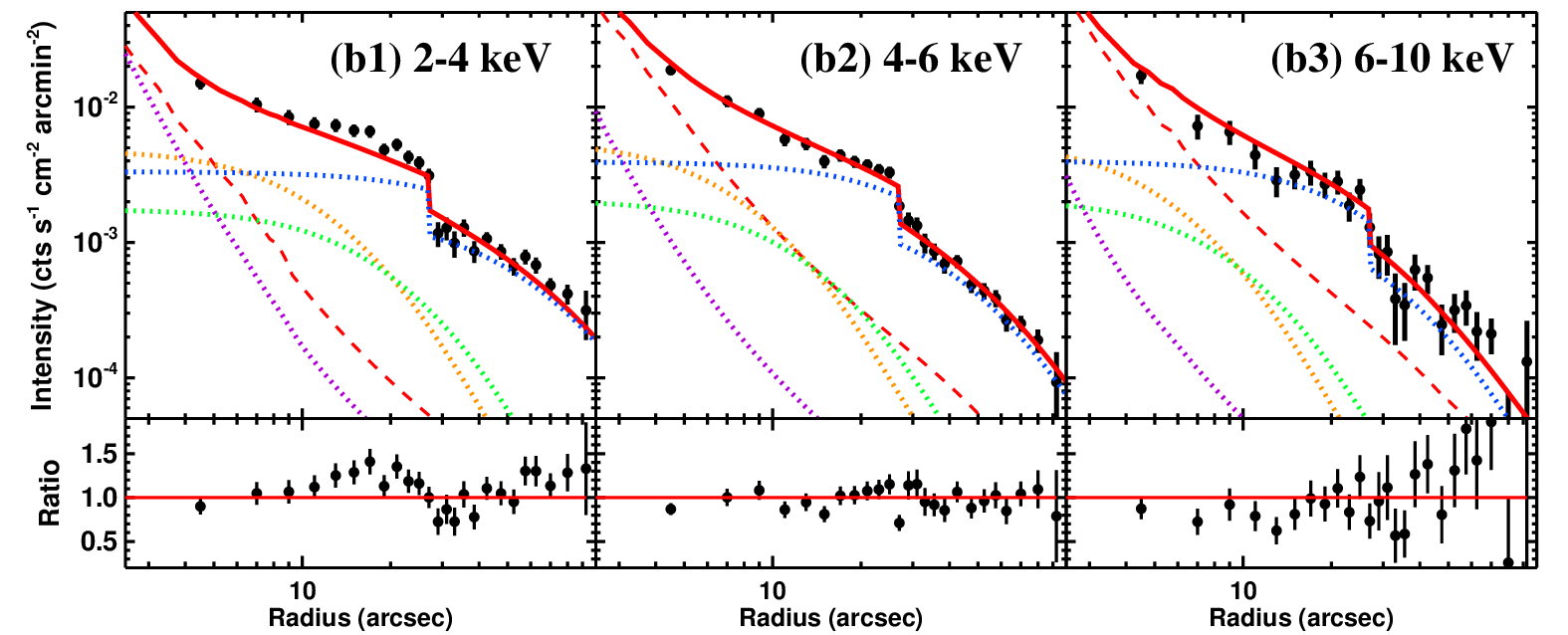}\\
\end{tabular}
\caption{Fitting the azimuthal-averaged halo profiles (Panels a1, a2 and a3) and the profiles in region-4 (Panels b1, b2 and b3), simultaneously, assuming an inhomogeneous dust distribution in layer-4 of Scenario-4.}
\label{fig-assyfit}
\end{figure*}

\section{Halo Azimuthal Asymmetry Analysis}
\label{sec-assy}
\subsection{Comparison of Halo Profiles in Subregions}
Figure~\ref{fig-region} shows that there is apparent azimuthal asymmetry in the halo. To investigate this, the region surrounding \swf\ was divided evenly into 6 sub-regions, each having an opening angle of 60 degrees. The 7th region was defined with a 30-degree angular width where the non-uniformity of the halo is most significant. Then a radial profile was extracted from each subregion. A flat background was subtracted in the same way as described in Section~\ref{sec-haloshape}. Figure~\ref{fig-assy-radp} shows radial profiles in every sub-region. The halo non-uniformity is most severe in region-3, 4, 7. Region-3 shows lower halo intensity than the azimuthal-averaged halo within 20-40 arcsec radii. Region-4 shows higher intensity within 10-30 arcsec radii. Region-7 is a sub-region of region-4, where the halo intensity drops sharply by a factor of 2-3 at the viewing angle of $\sim$ 30 arcsec.

To quantify this azimuthal asymmetry, we used the best-fit halo model from Scenario-4 with the COMP-AC-S grain population to fit the radial profile in every sub-region, by fixing the location of each layer but allowing the $N_{\rm H,sca}$ of layers-2 and 3 to vary. We also fixed the $N_{\rm H,sca}$ of layer-1 and layer-4, because they dominate the regions within 3 arcsec and outside 100 arcsec where the data quality is poor in sub-regions (see Figure~\ref{fig-halofit}). The best-fit $N_{\rm H,sca}$ in every subregion as a fraction of the azimuthal-averaged halo value is given in Table~\ref{tab-obs-assy}. We find that the $N_{\rm H,sca}$ of layer-2 can change by as much as a factor of 4 in different subregions, while the $N_{\rm H,sca}$ of layer-3 can change by a factor of 3, but the total $N_{\rm H,sca}$ in different subregions change by only $\sim$10\%.

However, simply changing the $N_{\rm H,sca}$ in layer-2 and 3 is not enough to account for all the halo asymmetry. For example, the flux drop in Figure~\ref{fig-assy-radp} Panel a7 at $\sim$30 arcsec is clearly too sharp to be fitted by any halo components. An obvious explanation is that the dust distribution is not uniform at all viewing angles in at least one of the foreground layers. This effect can potentially be associated with the distribution of molecular clouds in the source direction, which we discuss in more detail in Section~\ref{sec-discussion-assy}. 

\begin{table}
   \caption{Best-fit parameters from fitting the radial profiles in region-4 using Scenario-4, but assuming an inhomogeneous dust distribution in layer-4.}
    \label{tab-assyfit}
\begin{tabular}{lcccccc}
\hline
\hline
Layer & Parameter & Value & Unit \\
\hline
Layer-1 & x$_{\rm low,1}$ & 0.990$^{+0.002}_{-0.002}$ & \\
             & x$_{\rm high,1}$ & 1.000$^{u}_{-0.019}$ & & \\
             & $N_{\rm H,1}$ & 2.1$^{+0.2}_{-0.2}$ & $10^{22}$ cm$^{-2}$\\
\hline
Layer-2 & x$_{\rm low,2}$ & 0.857$^{+0.042}_{-0.042}$ & \\
             & x$_{\rm high,2}$ & 0.857$^{+0.104}_{-0.104}$ & \\
             & $N_{\rm H,2,ave}$ & 1.8$^{-0.5}_{-0.5}$ & $10^{22}$ cm$^{-2}$\\
             & $N_{\rm H,2,region-4}$ & 1.1$^{-0.2}_{-0.2}$ & $10^{22}$ cm$^{-2}$\\
\hline
Layer-3 & x$_{\rm low,3}$ & 0.765$^{+0.065}_{-0.065}$ & \\
             & x$_{\rm high,3}$ & 0.766$^{+0.065}_{-0.065}$ & \\
             & $N_{\rm H,3}$ & 1.8$^{-0.5}_{-0.5}$ & $10^{22}$ cm$^{-2}$\\
             & $N_{\rm H,3,region-4}$ & 1.0$^{+0.5}_{-0.5}$ & $10^{22}$ cm$^{-2}$\\
\hline
Layer-4 & x$_{\rm low,4}$ & 0.000$^{+0.075}_{l}$ & \\
             & x$_{\rm high,4}$ & 0.458$^{+0.090}_{-0.090}$ & \\
             & $N_{\rm H,4}$ & 10.5$^{+1.4}_{-1.4}$ & $10^{22}$ cm$^{-2}$\\
\hline
             & $N_{\rm H,4,region-4,inner}$ & 18.8$^{+1.8}_{-1.8}$ & $10^{22}$ cm$^{-2}$\\
             & $R_{\rm break}$ & 27.0$^{+0.1}_{-0.1}$ & arcsec \\
             & $N_{\rm H,4,region-4,outer}$ & 8.5$^{+0.8}_{-0.8}$ & $10^{22}$ cm$^{-2}$\\
\hline
\multicolumn{2}{l}{$\chi^2_{\rm \nu, ave}$ = 110.5/87}  & \multicolumn{2}{l}{$\chi^2_{\rm \nu, region-4}$ = 111.1/70} \\
\hline
\end{tabular}
\\
\\
{\bf Notes.} Error bars indicate the 1$\sigma$ parabolic errors. See Figure~\ref{fig-assyfit} for the corresponding halo decomposition.
\end{table}

\subsection{Halo Model with Inhomogeneous Dust Distribution}
In order to understand the flux drop in region-7, we created a toy model based on Scenario-4. From previous results, layer-4 has the largest contribution at large radii, so we assumed that its $N_{\rm H,sca}$ has a sudden change at a specific radius ($R_{\rm break}$, a free parameter) in region-7, then the sharp flux drop can be reproduced. We also allowed the $N_{\rm H,sca}$ of layers-2, 3 and 4 in region-7 to be different from the azimuthal-averaged values. Then a simultaneous halo profile fitting to the azimuthal-averaged halo and the one in region-7 was performed, and the results are shown in Figures~\ref{fig-assyfit} and Table~\ref{tab-assyfit}.

Firstly, the best-fit values from fitting the azimuthal-averaged halo and the halo in region-7 are very similar to the results of fitting only the azimuthal-averaged halo, only that the fraction of $N_{\rm H,sca}$ in layer-4 increases slightly to (64.8$\pm$8.6)\%, and its upper limit increases to 0.458$\pm$0.090. Secondly, the flux drop can be well modelled with $R_{\rm break}=27.0\pm0.1$ arcsec. Within this radius, the $N_{\rm H,sca}$ of layer-4 is ($18.8\pm1.8$)$\times$10$^{22}$ cm$^{-2}$ which is about 1.8 times as high as the average value. But it drops significantly to ($8.5\pm0.8$)$\times$10$^{22}$ outside $R_{\rm break}$, as required by fitting the sharp flux drop. The residuals in region-7, e.g. seen in Figure~\ref{fig-assyfit} Panel b1, is likely caused by more detailed radial variation of $N_{\rm H,sca}$ in foreground dust layers.

\section{Discussion}
\subsection{ISM Distribution along the Source Sightline}
\label{sec-discussion1}
\subsubsection{Results from the Halo Profile Study}
\begin{figure}
\includegraphics[trim=0.5in 1.8in 0in 0in, clip=1, bb=5 5 612 755, scale=0.43]{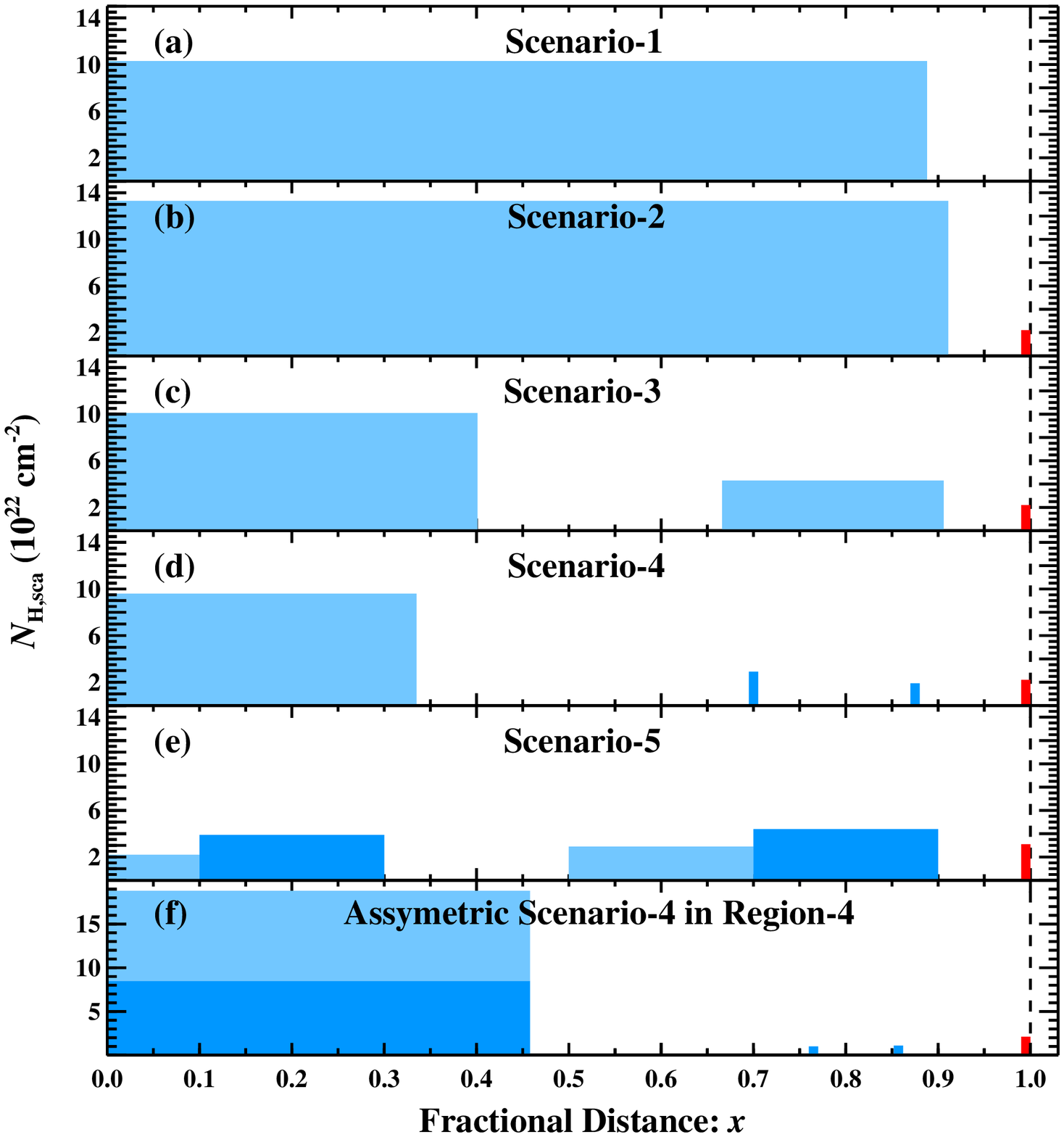}
\caption{The distribution of $N_{\rm H,sca}$, which is used as an indication of dust mass, along the source LOS as revealed by the best-fit models of different scenarios (Tables~\ref{tab-halofit},~\ref{tab-assyfit}). The dash line at $x = 1.0$ indicates the position of the source. The red histogram indicates that there is a local dust layer to the source. In Panel-f, the $N_{\rm H,sca}$ of the layer below $x < 0.5$ has a dramatic change from $8.5\times10^{22}$ cm$^{-2}$ inside the viewing angle of 27.0 arcsec (lower dark blue region) to the outside $18.8\times10^{22}$ cm$^{-2}$ (upper light blue region).}
\label{fig-dustdist}
\end{figure}

\swf\ is a heavily absorbed X-ray transient with $N_{\rm H,abs}~\simeq~2~\times10^{23}$ cm$^{-2}$ during the persistent phase (i.e. the time period out of dips, \citealt{Xu.2018}). We discover a strong X-ray dust scattering halo around it, which can be decomposed into several dust scattering components. Figure~\ref{fig-dustdist} shows the amount of dust, as indicated by the $N_{\rm H,sca}$, in every layer along the LOS from different best-fit halo models. It is clear that most of the X-ray scattering dust should be in the foreground layers far from \swf. We also tried some other dust grain populations, but found that they only had a small impact on the resultant relative dust distribution.

However, a potential uncertainty is associated with the fact that the halo profile modelling cannot fully constrain the intensity of the dust scattering component local to the binary system. This is because as the dust layer becomes closer to the primary X-ray source, the halo size also becomes smaller, eventually its size can be smaller than the pixel size if the dust layer is close enough to the X-ray source. In this extreme case, the source plus the local halo would still appear point-like, and the radial profile is undistinguishable from the instrumental PSF, i.e. the halo component is not `visible' in the observed radial profile even if the $N_{\rm H,sca}$ is large. Therefore, it is possible for the fraction of $N_{\rm H,sca}$ in the other foreground dust layers to be over-estimated.

\subsubsection{Results from the Molecular Cloud Study}
\label{sec-mcdist}
The distribution of molecular clouds (MCs) can reflect the distribution of ISM in the LOS of \swf. To determine the number of velocity-coherent components along the LOS, we use $^{13}$CO(1-0) data from the Mopra Southern Galactic Plane CO survey (\citealt{Burton.2013}). The spectrum is integrated over the circular region of 1 arcmin radius around \swf. To increase the S/N ratio, we also smoothed the spectrum, reaching an effective velocity resolution of 1.2 km s$^{-1}$. The velocity resolution is adequate for identifying different MC components.

\begin{figure*}
\centering
  \includegraphics[trim=0in 0.1in 0in 0in, clip=1, scale=0.6]{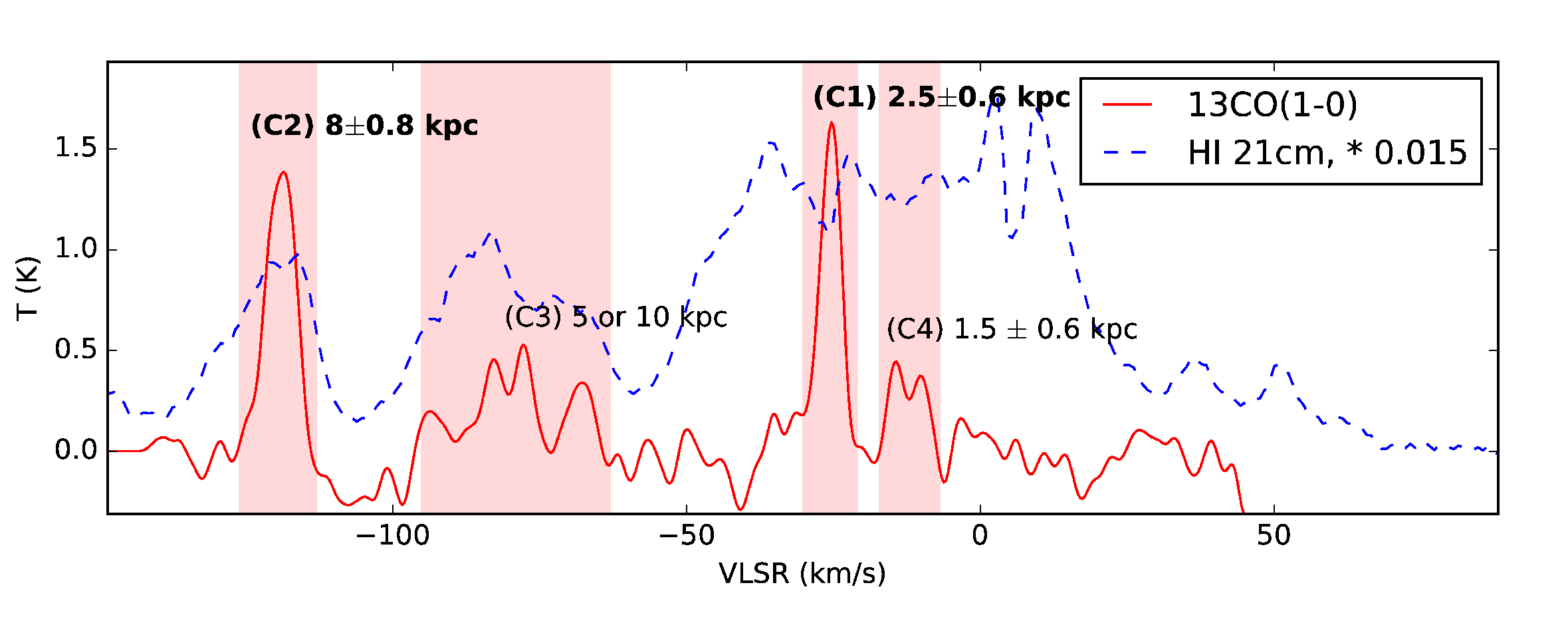}
  \caption{$^{13}$CO(1-0) and H {\sc i} 21cm spectra integrated over the region of 1 arcmin radius around \swf. The red  solid line is the $^{13}$CO(1-0) and the blue dashed line is the H {\sc i} 21cm spectrum. The identified components and their distances are indicated. See Section~\ref{sec-mcdist} for details. VLSR: velocity of observer relative to local standard of rest.}
  \label{fig-mcdist}
\end{figure*}

The $^{13}$CO(1-0) spectrum is plotted in Figure~\ref{fig-mcdist}. Firstly, we identified components and measured their velocities. The distances of these components can be determined based on their velocities using the Galactic model described in \citet{Reid.2016}, where their displacements from the plane, and proximities to individual parallax sources are automatically considered. We identified four components and estimated their distances. These are indicated in Figure~\ref{fig-mcdist}, and are summarized as follows:
\begin{itemize}
\itemsep0em
\item A intensive component (C1) at -25 km s$^{-1}$, which sits at a distance of $2.5\pm 0.6$ kpc.
\item A intensive component (C2) at -118 km s$^{-1}$, which sits at a distance of 8.0 $\pm$ 0.8 kpc.
\item A weaker component (C3) at -11 km s$^{-1}$, with a relatively low surface densities, which sits at a distance of 1.5 $\pm$ 0.6 kpc.
\item A component (C4) at -72 km s$^{-1}$, with a relatively large velocity dispersion, which sits at a distance of either 5 or 10 kpc.
\end{itemize}
C1 and C2 have integrated intensities of $\sim$12 K km s$^{-1}$ and 24 K km s$^{-1}$ respectively. Using the standard conversion in \citet{Simon.2001}, these correspond to column densities of $1.2\times10^{22}$ cm$^{-2}$ and  $2.4\times10^{22}$ cm$^{-2}$, respectively. Note that these are the mean column densities estimated by integrating over a large region, and they are not necessarily identical to the column density derived using dust scattering. We also note that C1 and C2 have well-defined emission peaks, and they probably correspond to dense MCs in the Galactic disk. C3 is much weaker might correspond to a cloud of a lower surface density. C4 consists of several peaks and is wide in velocity (${\rm d} v \approx$ 20 km s$^{-1}$). It probably corresponds to diffuse gas that lingers around in one of the the Galactic spiral arms.

Conventionally, the distance of a Galactic source can be determined using its velocity and its H {\sc i} 21 cm spectrum, where its distance can be determined using its velocity with the help of a Galactic rotation model. In most cases, distance estimated using the Galactic rotation method has two solutions, where the H {\sc i} 21 cm emission line can be used to distinguish between these two solutions, e.g. if a source stays close to us, it can produce an absorption feature visible in the spectrum. In this work, our distances are mostly determined with the help of  the Galactic model of \citet{Reid.2016} where the H {\sc i} 21 cm data is not used. Although this is already considered as accurate, we still plotted the H {\sc i} 21 cm emission presented in \citet{McClure-Griffiths.2005} as a reference, where C1 and C3 do exhibit corresponding absorption features, implying that they presumably stay at the near distances. Components C2 and C4 show very little self-absorption indicating that they probably stay at the far distances. These are broadly consistent with the result from the galactic model described in \citet{Reid.2016}. If the source is located at a distance of $\sim$10 kpc (see Section~\ref{sec-discussion-distance}), then most of the MCs in its LOS will be located in the Galactic disk.

\begin{figure*}
\centering
\includegraphics[bb=460 0 750 610, clip=1, scale=0.42]{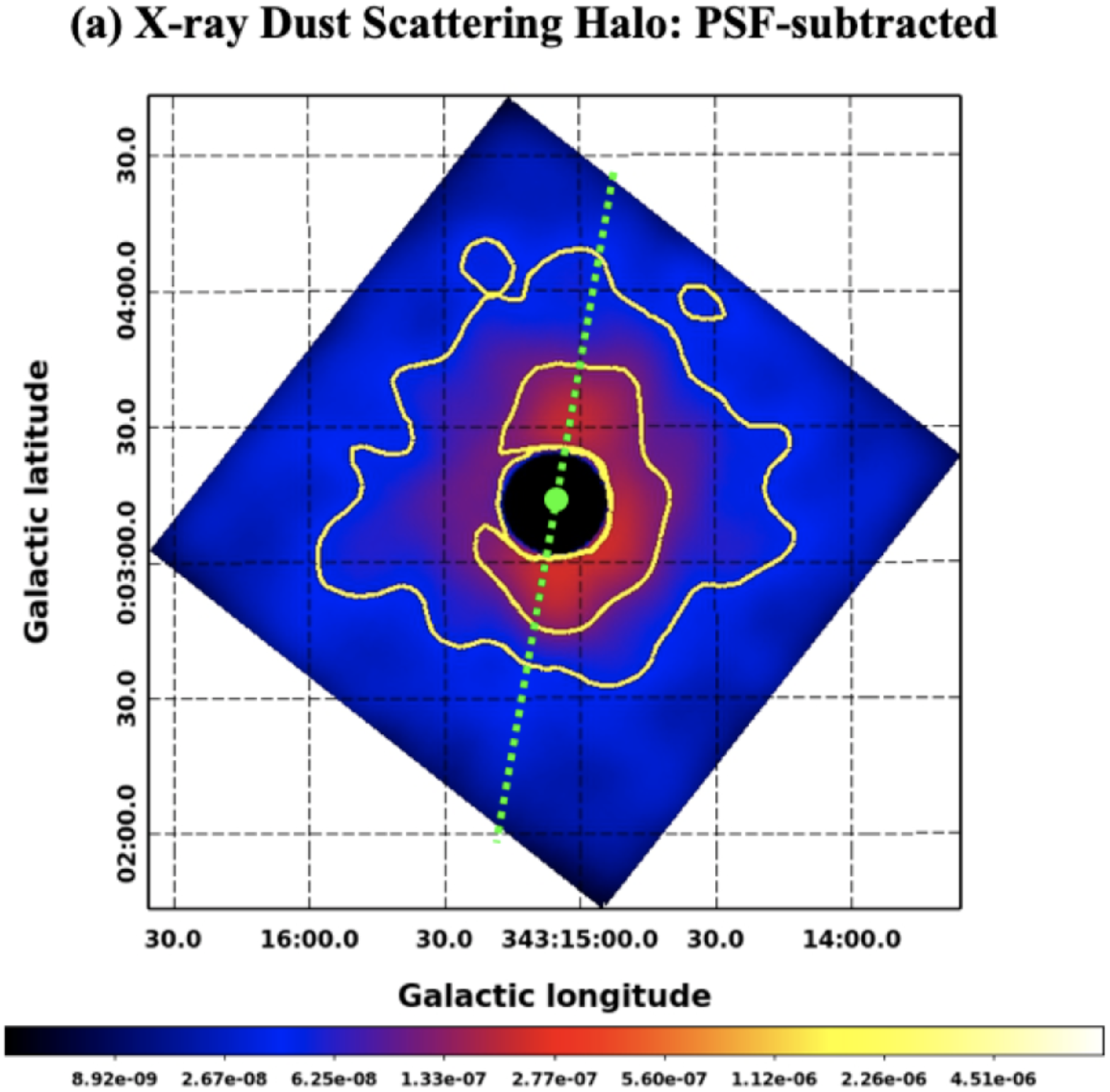}
\caption{Panel-a: the smoothed 4-6 keV dust scattering halo image around \swf, after subtracting the background, readout streak and the PSF. The yellow contours reflect the intensity distribution.  The green point indicates the source position. The green dotted line indicates the position of the removed readout streak. Panel-b: the ATLASGAL 0.87mm image of the 1 arcmin region around \swf. The yellow contours are directly copied from Panel-a to compare the intensity distribution of dust scattering halo with the 0.87mm emission from molecular clouds.}
\label{fig-assy-image}
\end{figure*}

\subsubsection{More Evidences for the ISM Distribution}
There are other clues to support the fact that a large fraction of the gas and dust along the LOS of \swf\ should be located in the Galactic disk, which are described separately below:
\begin{itemize}
\itemsep0em
\item The total $N_{\rm H,sca}$ of \swf\ is found to be $(1.66\pm0.16)\times10^{23}$ cm$^{-2}$ (see Table~\ref{tab-halofit}, Scenario-3). The dust grain model adopts the Solar abundances from \citet{Holweger.2001}, which is also consistent with the \citet{Asplund.2009} abundances. In comparison, the $N_{\rm H,abs}=(1.8\pm0.08)~\times10^{23}$ cm$^{-2}$ for the persistent phase outside the dips\footnote{The $N_{\rm H,abs}$ of \swf\ can reach $10^{24}$ cm$^{-2}$ during the dips (\citealt{Xu.2018}). In this case most of the X-ray absorption should be intrinsic to the binary system, likely due to the matter transferred from the companion star.} (\citealt{Xu.2018}) is based on the interstellar medium (ISM) abundances from \citet{Wilms.2000}, which is $\sim20$\% lower than the Solar abundances. Then $N_{\rm H,abs}$ based on the \citet{Holweger.2001} Solar abundances will be $(1.99\pm0.19)~\times10^{23}$ cm$^{-2}$. Thus $N_{\rm H,abs}$ is consistent with $N_{\rm H,sca}$, which implies that the detected LOS dust in the halo can already fully account for the observed X-ray absorption.
\item The $N_{\rm H,abs}$ of \swf\ has a baseline level on the order of $10^{23}$ cm$^{-2}$ in all the observations taken at different times and different source flux levels (Bogensberger et al. in preparation). This underlying $N_{\rm H,abs}$ is most likely to be associated with the ISM in the Galactic disk, which should not vary as rapid as the ISM local to the binary system of \swf\ since its discovery.
\item According to the \citet{Schlafly.2011} reddening map, the Galactic extinction in the LOS of \swf\ is $A_{\rm v}\sim58.6$ mag (\citealt{Pessev.2018}), which, using the \citet{Zhu.2017} gas-to-dust relation, corresponds to $N_{\rm H,abs}~=~1.22~\times10^{23}$ cm$^{-2}$ (\citealt{RussellDM.2018}). This value is also consistent with the observed lower limit of $N_{\rm H,abs}$. Thus the strong Galactic reddening towards \swf\ also implies a large fraction of foreground gas and dust in the Galactic disk.
\item \swf\ is in the Galactic plane at 16.693 degree from \sgra, its underlying $N_{\rm H,abs}$ is very similar to \sgra\ (\citealt{Ponti.2017}) and the GC magnetar SGR J1745-2900 (e.g. \citealt{Kennea.2013}; \citealt{Mori.2013}; \citealt{Rea.2013}; \citealt{CotiZelati.2017}; \citealt{Jin.2017}). It has been reported that most of the GC foreground gas and dust should be located in the Galactic disk, probably distributed along the spiral arms (e.g. \citealt{Bower.2014}; \citealt{Wucknitz.2015}; \citealt{Sicheneder.2017}; \citealt{Jin.2017,Jin.2018}). Thus it is possible that the same intervening spiral arms also create significant X-ray absorption opacity and dust scattering opacity towards \swf.
\end{itemize}

According to all the above clues, we conclude that most of the gas and dust in the LOS of \swf\ is likely to be located in the Galactic disk, and they correspond to an $N_{\rm H}$ of the order of $10^{23}$ cm$^{-2}$.

\subsection{Asymmetric Halo and Inhomogeneous Distribution of Foreground ISM}
\label{sec-discussion-assy}
The azimuthal asymmetry of the X-ray dust scattering halo has been reported in some sources (e.g. \citealt{Seward.2013}; \citealt{Valencic.2015}). The halo can become non-uniform because of at least two reasons. One reason is the partial alignment of non-spherical dust grains with the magnetic field. This type of halo azimuthal asymmetry can reach 10\% at the same radius, as calculated by \citet{Draine.2006}. Another reason is the non-uniform distribution of foreground dust, which can cause different scattering opacity at different viewing angles and azimuthal angles. The halo asymmetry observed in \swf\ changed by a factor of 2-3 at the radius of 30 arcsec in region-7, as shown in Figure~\ref{fig-region}, which is much larger than the asymmetry that can be caused by the magnetic field. The change of dust grain population cannot explain the asymmetry, because the profile of any dust scattering halo cannot reproduce the sharp flux drop in region-7. Thus the most likely explanation is the inhomogeneous distribution of foreground dust. Indeed, in Section~\ref{sec-assy} we have shown that a sudden $N_{\rm H,sca}$ drop at the viewing angle of 27 arcsec in layer-4 can account for most of the observed flux drop in region-7.

The 2-dimensional (2D) azimuthal asymmetry of the halo can also be visualized. To do this, we first reran the {\tt acisreadcorr} script to remove the readout streak but keeping a set of photons whose spectrum is similar to pixels at the same radius{\footnote{\url{http://cxc.harvard.edu/ciao/ahelp/acisreadcorr.html}}}, so that the region of the readout streak was no longer black after the removal. Note that this step is only for the purpose of creating nicer images of the halo, because it is not possible to fully recover the primary halo intensity underneath the readout streak. Then we renormalized the {\sc ChaRT}-simulated 4-6 keV PSF image to the intensity of the re-extracted 4-6 keV flux image within the 2-4 arcsec annulus, and subtracted the PSF image from the flux image, leaving only the dust scattering halo. Figure~\ref{fig-assy-image}-a shows the smoothed halo image, where complex azimuthal inhomogeneity of the halo is clearly revealed by yellow contours.

Since the dust distribution in the ISM is likely to be linked to the distribution of MCs, it is possible to find this inhomogeneity in the MC distribution in the same direction. Figure~\ref{fig-assy-image}-b shows the 0.87 mm intensity map, which is an indicator for the intensity map of the CO 3$\rightarrow$2 emission line, from the Atacama Pathfinder EXperiment (APEX) Telescope's large Area Survey of the Galaxy (ATLASGAL) (\citealt{Schuller.2009}) in the direction of \swf. The yellow contours from the halo intensity image is copied to the ATLASGAL image for comparison. This shows that there is an evident correlation between the X-ray halo intensity and the intensity of the 0.87 mm emission. Therefore, it is most likely that the asymmetric X-ray dust scattering halo is due to the inhomogeneous distribution of foreground ISM.

\subsection{Distance to \swf}
\label{sec-discussion-distance}
\swf\ is a heavily extincted source in the Galactic plane. The strong dust reddening and X-ray absorption require the source to be far from Earth, so that its LOS can pass through enough ISM in the Galactic disk to produce such severe extinction.
Soon after the discovery of \swf, \citet{RussellTD.2018} compared its radio and X-ray luminosities, and suggested that at a distance of $d$ > 3 kpc the source appear like a black hole X-ray binary, while at closer distances it would be more consistent with a neutron star X-ray binary. Then the following study of \citet{Xu.2018} showed that the spectral timing properties of \swf\ look like a black hole binary, which seems to support $d$ > 3 kpc. However, it is not possible to get better constraints on the source distance based on this argument alone. Below we discuss several other methods of estimating the source distance.

\subsubsection{Distance Estimated from the State Transition Luminosity}
An independent method to estimate the source distance is to use the soft-to-hard state transition Eddington ratio, i.e. the transition luminosity in the unit of Eddington luminosity. \citet{Maccarone.2003} studied the transition luminosity of 10 X-ray binaries (XRBs), including 6 black hole XRBs and 4 neutron-star XRBs. They found that for black hole XRBs the soft to hard transition luminosity is typically within 1-4\% of the Eddington Luminosity, with an average value of 2\%; while for neutron-star XRBs it is within 0.4-5\%. Thus the transition luminosity can be used as a rough estimate of the source distance, provided that the mass of the accreting source is known.

Following the hardness ratio evolution of \swf\ as monitored by {\it Swift} satellite, the soft to hard transition flux in the 2-10 keV band was observed to be $\sim 1.5\times10^{-10}$ erg s$^{-1}$ cm$^{-2}$ (Bogensberger et al. in preparation). Then the bolometric transition flux from 0.5 keV to 10 MeV can be estimated to be $\sim 1.4\times10^{-9}$ erg s$^{-1}$ cm$^{-2}$, assuming a power law model with the photon index 1.8, $N_{\rm H,abs}=1.66\times10^{23}$ cm$^{-2}$ and an exponential cutoff at 200 keV (\citealt{Maccarone.2003}). If \swf\ contains a 1.5 M$_{\odot}$ neutron star, then for the transition Eddington ratio of 0.4-5\%, $d$ would be 2.2-7.6 kpc. However, \swf\ is more likely to be a black hole XRB (\citealt{Xu.2018}). Then for a black hole mass of 5-15 M$_{\odot}$, and for the average transition Eddington ratio of 2\%, $d$ would be 8.8-15.3 kpc. We can also calculate that $d$ changes to 6.2-10.8 kpc and 10.8-18.7 kpc for the transition Eddington ratio of 1\% and 4\%, separately.

\subsubsection{Distance Estimated from the Halo Profile}
\citet{Valencic.2015} used a single foreground dust layer to fit the halo profiles of 35 X-ray sources with a large range of LOSs. They discovered that for X-ray sources whose LOSs traversed more than 5 kpc in the Galactic plane, their halo profiles could not be well fitted by a single foreground dust layer. \swf\ also requires at least 3 dust layers to fit its halo, so it is likely to have $d>$ 5 kpc. However, since all the distances in the model are fractional, and they are also degenerated to the dust grain population to some extent (see Tables~\ref{tab-halofit}, \ref{app-tab-halofit1}, \ref{app-tab-halofit2}), it is not possible to obtain further constraints on the source distance using only the halo profile study. The fractional distances must be mapped to the real distribution of ISM in order to determine the source distance more precisely.

\subsubsection{Distance Estimated from the LOS MC Distribution}
Since a large fraction of interstellar dust is likely to be contained in MCs, the distribution of MCs in the LOS can be used to compare with different dust scattering components, and thus help to determine the source distance. In Section~\ref{sec-mcdist} we have shown that there are at least 4 major MC components in the LOS of \swf. Comparing with the best-fit dust distribution in Scenarios-3, 4 and 5, C2 can be associated with the layers within $0.7 \le x \le 0.9$ where $\sim$25\% LOS dust is located. This allows us to estimate a source distance of 9 $\lesssim d \lesssim$ 11 kpc. This distance also put C1 and C3 in the fractional distance range of $0.0 \le x \le 0.4$ where $\sim$50\% LOS dust is located. Moreover, if C4 is located at 10 kpc away and is associated with layer-1 which is intrinsic to \swf, then the source distance will also be 10 kpc.

Considering all the independent distance estimates discussed above, we conclude that \swf\ is most likely to be located at $d \sim$ 10 kpc.

\begin{figure}
\begin{tabular}{c}
\includegraphics[clip=1, bb=70 200 558 625, scale=0.475]{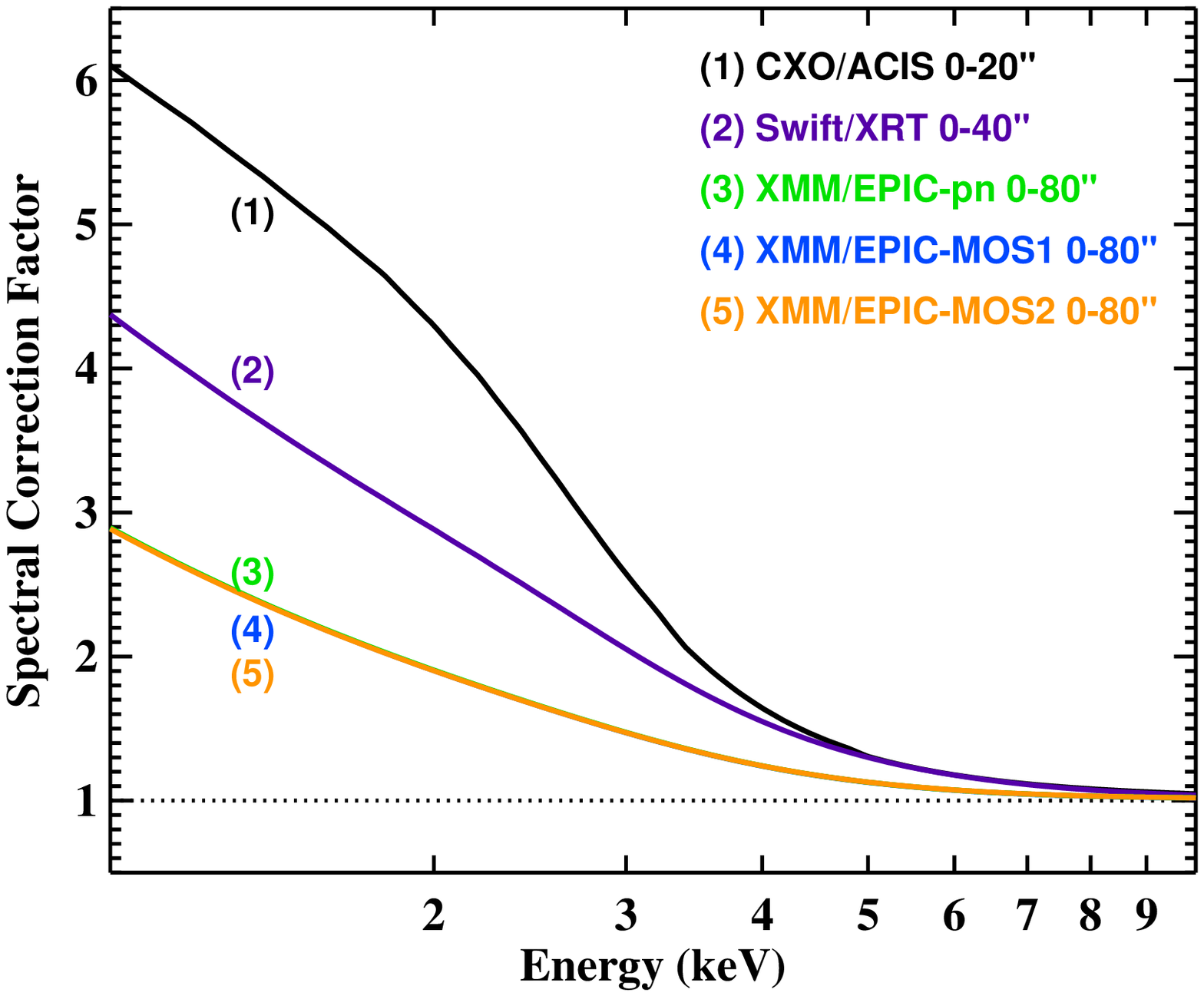} \\
\includegraphics[clip=1, bb=54 216 558 490, scale=0.475]{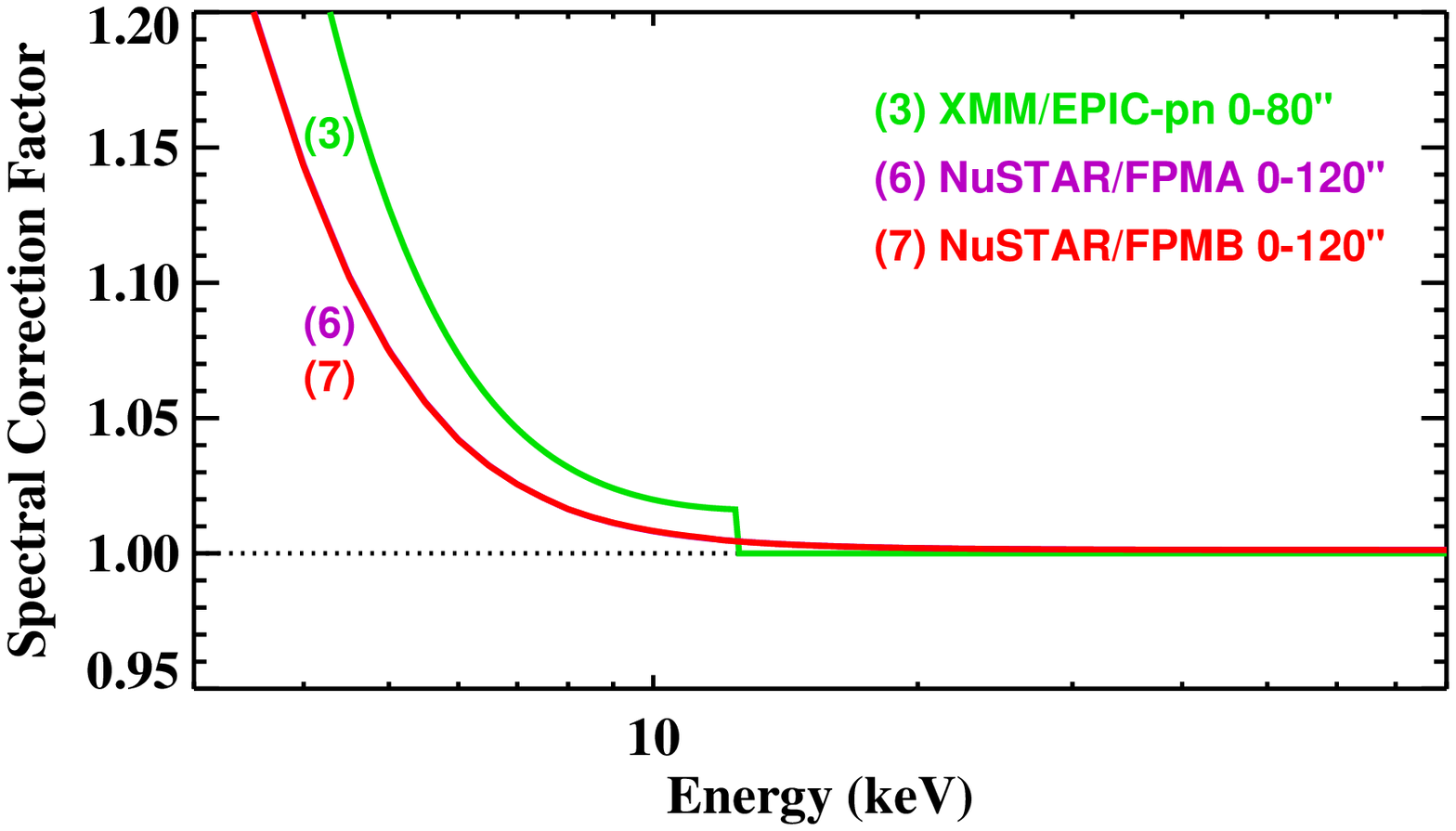} \\
\end{tabular}
\caption{The predicted spectral correction factor, as a function of energy, for typical source extraction regions in different instruments, based on the best-fit halo model in Scenario-4 using the COMP-AC-S grain population, as shown in the Panels 4a, 4b and 4c in Figure~\ref{fig-halofit}. The factor is fixed at 1.0 for XMM/EPIC-pn after 12.0 keV because it does not cover higher energy bands.}
\label{fig-specratio}
\end{figure}

\subsection{Spectral Bias Caused by the Dust Scattering Halo}
\label{sec-discussion-bias}
The strong dust scattering halo around \swf\ makes its radial profile different from the PSF profile in any instrument, thus the normal PSF correction used to retrieve the intrinsic source flux is not accurate unless the flux in the halo is properly treated (\citealt{Smith.2016}; \citealt{Jin.2017}). As the photon energy decreases, the halo becomes more significant, and so the spectral bias caused by the halo also increases. Due to the PSF convolution, the spectral bias is also more significant for instruments with smaller PSF. The source extraction region also affects the level of the spectral bias. If the region covers the entire halo, then the loss of photons in the LOS due to the dust scattering opacity is fully compensated by the photons in the halo, and so there will be no spectral bias. However, a typical source extraction region has a limited size, and part of the central region is often excluded to avoid the pile-up effect, thus the spectral bias induced by the halo is inevitable.

The correction factor required to remove the spectral bias can be calculated from the halo model at different energies for different source extraction regions and instrumental PSF profiles. We used the best-fit halo model in Scenario-4 to calculate the correction, as this model gives the smallest $\chi^2$. It must be emphasized that an accurate correction factor only requires the halo model to fit the halo shape accurately in every energy band, it does not depend on the detailed halo decomposition. This is because the shapes of all the halo components, except the PSF, have the same energy dependence (e.g. \citealt{Mathis.1991}; \citealt{Predehl.1995}). Therefore, all the uncertainties and model degeneracies associated with the best-fit halo parameters do not affect the spectral correction (e.g. see Fig.11 in \citealt{Jin.2017}).

Figure~\ref{fig-specratio} shows the correction factor as a function of energy for typical source extraction regions in different instruments. It is clear that instruments with smaller PSF need larger correction factors, and the factor increases rapidly as the energy decreases. Based on these model calculations, we create {\sc xspec} models, named {\tt dscor}, for different instruments and different source extraction regions{\footnote{The model package will be made available at \url{http://blog.jinchichuan.cn/dscor-model} and submitted to the {\sc xspec} website. It can also be downloaded directly from \url{https://www.dropbox.com/s/mvpvvas0i9y3hcs/dscor.tar.gz?dl=0}.}}.

Another method often used to correct for the spectral bias of dust scattering is to assume a gas-to-dust ratio and add the dust scattering opacity to the X-ray absorption model. This method requires that the halo intensity should be negligible compared to the PSF intensity in the source extraction region. Firstly, this requirement is difficult to be met for heavily absorbed X-ray sources like \swf\ and GC sources (Jin et al. in preparation), because their dust scattering opacity is not negligible. Secondly, in the case of lower X-ray absorption, if the dust is intrinsic to the source, the halo will have the same profile as the PSF (\citealt{Mathis.1991}), then the effective dust scattering opacity in the extracted spectrum would be zero, i.e. the dust is not `visible' to the observer (\citealt{Jin.2017}), and so there would be no need to consider it in the absorption model. Therefore, this alternative spectral correction method is problematic. {\it We emphasize that it is a must to measure and model the profile of the dust scattering halo in order to properly correct for its spectral biases.}

\section{Summary and Conclusions}
\swf\ is a bright X-ray transient in the Galactic plane at 16.693 degrees away from \sgra. Its X-ray emission was heavily absorbed with $N_{\rm H, abs}\sim2\times10^{23}$ cm$^{-2}$. Given the X-ray brightness and the large amount of LOS dust as inferred from $N_{\rm H, abs}$, we carry out a detailed study of the X-ray dust scattering halo around it. The main results of this work are summarized below:
\begin{itemize}
\itemsep0em
\item We discovered a strong X-ray dust scattering halo around \swf\ in the data of the latest \cxo\ and \xmm\ observations.
\item The azimuthal-averaged halo was well modelled by at least 3 dust layers along the LOS. The best-fit halo model with the COMP-AC-S dust grain population had a total $N_{\rm H, sca}$ which is consistent with the $N_{\rm H, abs}$ of the X-ray spectrum observed during the persistent emission phase. Our halo modelling also suggested that 85-90 percent gas and dust in the LOS of \swf\ should be located in the Galactic disk far from the source.
\item The halo showed significant azimuthal asymmetry, with one of the azimuthal sub-regions exhibiting a sharp flux drop in the halo profile. This feature was well explained as a sudden change of dust column in the major halo component nearby at the viewing angle of $27.0\pm0.1$ arcsec.
\item We found that the 2D halo shape is broadly correlated with the 2D intensity map of CO ($3\rightarrow2$) 0.87 mm emission, suggesting that the halo asymmetry should be due to the inhomogeneous distribution of dust in the foreground ISM.
\item Several methods were used to obtain independent estimates of the source distance. The most direct constraints came from the comparison between the distributions of halo components and MC components. Our results indicated that \swf\ is most likely to be at $d\sim$ 10 kpc.
\item The strong dust scattering halo introduced strong spectral biases for different instruments with conventional source extraction regions. We produced {\sc xspec} models to correct for the spectral biases.
\item We explained why it is problematic to simply assume a gas-to-dust ratio to estimate and correct for the dust scattering opacity. Instead, it is a must to measure and model the halo profile in order to correct for it properly. 
\end{itemize}

The results of our work also highlighted the importance of considering the dust scattering in other absorbed X-ray sources, especially those with $N_{\rm H, abs} \gtrsim 10^{23}$ cm$^{-2}$. Since the halo shapes of different sources are very likely to be different, this spectral correction must be carried out on a source-by-source basis in order to achieve the highest accuracy.

\acknowledgments

We thank the anonymous referee for providing valuable comments and suggestions. CJ thanks Ligang Hou for insightful discussions about the molecular clouds in the Milky Way. GL thanks Xiaohui Sun for help with the 21cm data. The authors wish to thank the \cxo\ and \xmm\ teams for approving and scheduling the DDT observations. CJ acknowledges the National Natural Science Foundation of China through grant 11873054. GP acknowledges financial support from the Bundesministerium fur Wirtschaft und Technologie/Deutsches Zentrum f\"{u}r Luft- und Raumfahrt (BMWI/DLR, FKZ 50 OR 1604, FKZ 50 OR 1715, FKZ 50 OR 1812) and the Max Planck Society. This work is mainly based on observations made by the \cxo\ X-ray Observatory, as well as data obtained from the \cxo\ Data Archive. We also used the data from \xmm, an ESA science mission with instruments and contributions directly funded by ESA Member States and NASA. This research has made use of software provided by the \cxo\ X-ray Center (CXC) in the application packages CIAO, ChIPS, and Sherpa.

%

\vspace{5mm}
\facilities{Chandra(ACIS), XMM-Newton(EPIC)}


\software{
{\sc ciao} (v4.10; \citealt{Fruscione.2006}), 
{\sc sas} (v17.0.0; \citealt{Gabriel.2004}),  
{\sc esas} (\citealt{Snowden.2014}),
{\sc ds9} (v7.5; \citealt{Joye.2003}),
{\sc iminuit} (v1.3.3, \citealt{James.1975}),
{\sc xspec} (v12.10.1; \citealt{Arnaud.1996})
}





\appendix
\section{Dust Scattering Halo Observed by \xmm\ MOS}
\label{app-sec-mos}
\begin{figure*}
\includegraphics[trim=0in 0.3in 0in 0in, clip=1, bb=0 0 990 530, scale=0.5]{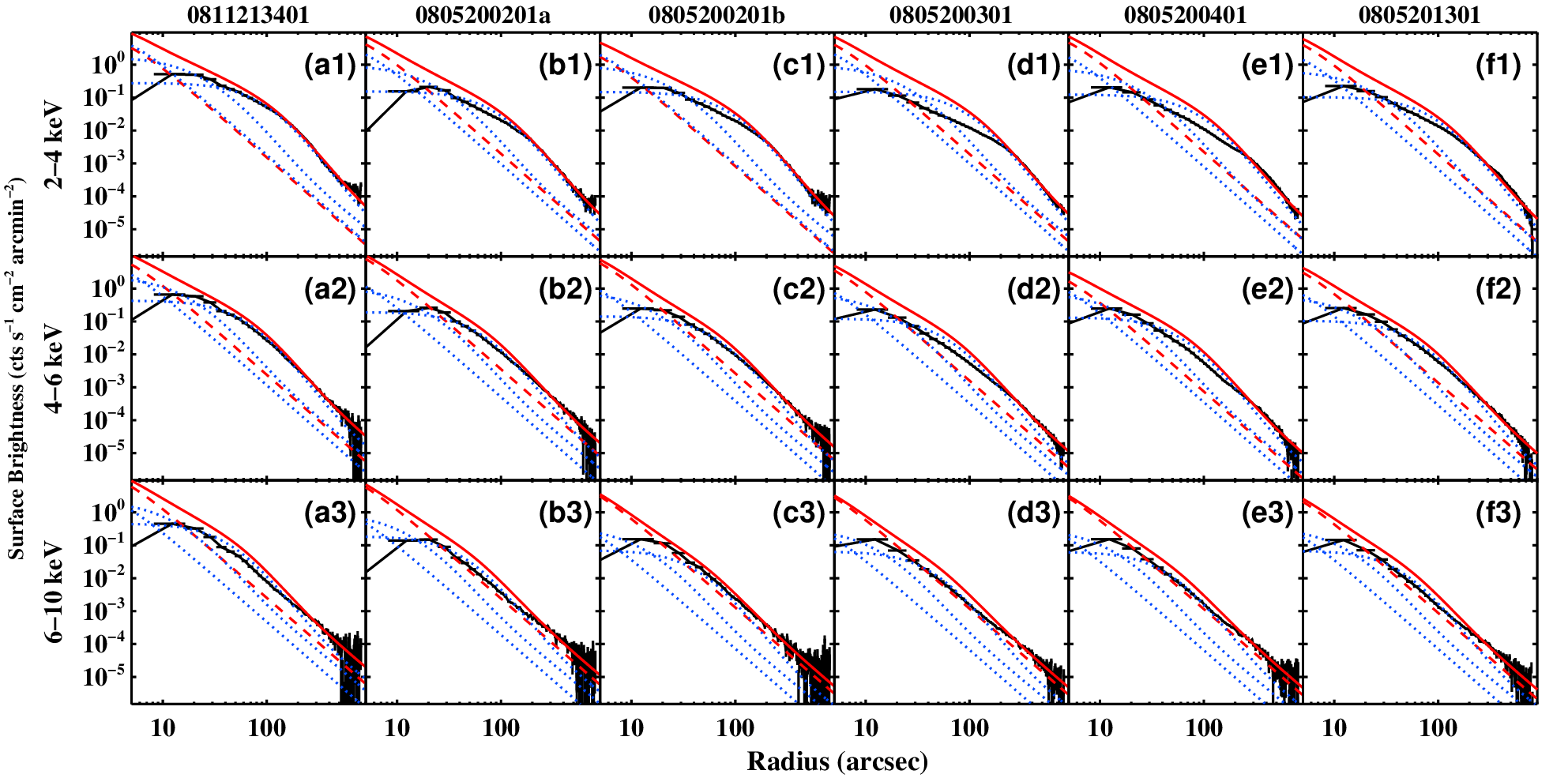}
\caption{The background subtracted radial profiles of \swf\ observed in \xmm\ MOS, as shown by the black points connected with the black solid line. In each panel, the red solid line is the best-fit halo model based on the \cxo\ data, convolved with the MOS PSF, renormalized to the flux of \swf\ in each \xmm\ observation outside 300 arcsec radius. The red solid line is the total halo model, the red dash line is the MOS PSF, all the orange dotted lines represent different halo components in Scenario-1 (as shown in the Panels a1, b1 and c1 in Figure~\ref{fig-halofit}). The discrepancy between the model and data at $\lesssim$ 200 arcsec radii is due to the severe pile-up effect.}
\label{app-fig-xmm-radp}
\end{figure*}

The halo profile observed by \xmm\ MOS should be the convolution of the intrinsic halo shape with the MOS PSF, as shown by \citet{Jin.2017} for the halo around \axj. We extracted the radial profiles of \swf\ from each of the time intervals as described in Section~\ref{sec-xmm} within 2-4, 4-6 and 6-10 keV bands. Since \swf\ was very bright during the \xmm\ observations, it caused severe pile-up effect on the CCD array, thereby distorting the halo shape in MOS. This effect is confirmed by the obvious bending-down of the radial profile at inner radii as show in Figure~\ref{app-fig-xmm-radp}. 

\begin{figure*}
\includegraphics[trim=0.95in 5in 0in 0in, clip=1, bb=0 0 990 560, scale=0.61]{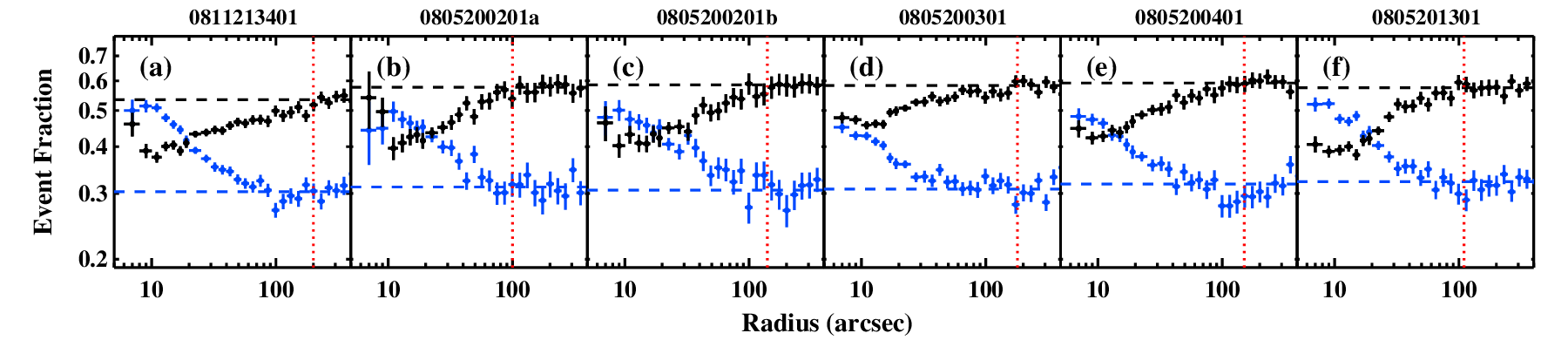}
\caption{The fraction of different events in 7-10 keV as a function of radius from \swf. In every panel, black points indicate the fraction of single events, while blue points indicate the fraction of double events. The horizontal dash lines indicate the pile-up-free fraction as determined from the data at large radii where the fraction is constant. The vertical red dotted line indicates the radius where the ratio of single events begins to deviate from the pile-up-free value.}
\label{app-fig-xmm-pileup}
\end{figure*}

In comparison, the photon pile-up in the \cxo\ ACIS observation is less than one percent outside 2.5 arcsec radius. It is possible for the dust layer local to \swf\ to change its $N_{\rm H, sca}$ significantly within a short timescale. For example, $N_{\rm H, abs}$ has been observed to increase by one order of magnitude during the dipping periods (\citealt{Xu.2018}). However, the timescale for the foreground dust layers in the Galactic disk to vary significantly may be much longer (see Section~\ref{sec-discussion1}). Therefore, we can convolve the best-fit halo model from Scenario-3 (see Section~\ref{sec-halofit}) with the PSF of MOS to obtain the predicted pile-up-free radial profile in MOS. Then we used the halo model, together with a free constant to account for the background, to fit the observed radial profile outside 300 arcsec where the halo should be dominated by dust scattering components far from the source. Then we extrapolated the model down to smaller radii. Figure~\ref{app-fig-xmm-radp} shows the background-subtracted radial profiles and the corresponding halo models. We find that the observed halo profile in MOS has a consistent shape with the model at $\gtrsim$ 200 arcsec. Within 200 arcsec, the observed profile is always lower than the model extrapolation, and the discrepancy increases towards smaller radii, which is consistent with the expected behaviour of photon pile-up effect.

In order to further check that the apparent differences between the observed and model radial profiles in MOS are not due to the change of $N_{\rm H,sca}$ in any foreground dust layers, we also show individual dust scattering components (blue dotted lines) and the PSF profile (red dash line) in Figure~\ref{app-fig-xmm-radp}. It is clear that the observed radial profile in MOS is flatter than any of the halo components, thus it cannot be fitted by any linear combination of these halo components, and so the change of $N_{\rm H,sca}$ cannot be the reason.

Since photon pile-up can reduce the fraction of single events and increase multiple-pixel events, we computed the fraction of single and double events in 7-10 keV. Figure~\ref{app-fig-xmm-pileup} shows the results for every observation. The pile-up-free fraction can be determined from the data at large radii where there is no pile-up and the fraction is constant, as shown by the dash lines in the figure. As the pile-up effect becomes stronger at inner radii, the fraction of single events becomes smaller, while the fraction of double events becomes larger. Therefore, the outer radius of the pile-up region can be roughly estimated to be the radius where the observed fraction of single events begins to deviate from the pile-up-free value, as shown by the red dotted line in the figure. We found that this radius is within the range of 100-200 arcsec, which is consistent with the radius where the observed halo shape deviates from the prediction. Because of the severe pile-up effect in MOS, we cannot extract further information from the halo profile.

\section{Modelling the Azimuthal-Averaged Halo Profile with Different Dust Grain Populations}
\label{app-sec-halofit}
For the five halo model scenarios described in Section~\ref{sec-halofit}, we changed the dust grain population to BARE-GR-B and COMP-NC-B, and repeated the above analysis. The best-fit parameters are listed in Tables~\ref{app-tab-halofit1} and \ref{app-tab-halofit2}. For each grain population, the $\chi^2$ improves significantly from one layer to two and three, but only marginally as the number of layers reaches 4. Scenario-5 produces worse fit than Scenario-3 and 4. Since the BARE-GR-B population is characterized by dust grains with smaller sizes than COMP-AC-S, its halo profile is more extended, therefore the best-fit fractional distances are all closer to the source. In comparison, the COMP-NC-B population has a bigger fraction of large dust grains, so the best-fit fractional distances are all closer to Earth. The relative distribution of LOS dust for each scenario does not change significantly with the dust grain population. The total $N_{\rm H,sca}$ is $(1.92\pm0.16)~\times10^{23}$ cm$^{-2}$ for BARE-GR-B and $(2.61\pm0.13)~\times10^{23}$ cm$^{-2}$ for COMP-NC-B in Scenario-3. These values are larger than in COMP-AC-S, which is due to different grain metallicities and size distributions assumed in different grain populations.

\begin{table*}
 \centering
  \begin{minipage}{175mm}
  \centering
\caption{Best-fit halo models in different scenarios using the BARE-GR-B dust grain population.}
   \label{app-tab-halofit1}
\begin{tabular}{lcccccccccc}
\hline
\hline
Layer & Parameter & Scenario-1 & Scenario-2 & Scenario-3 & Scenario-4 & Scenario-5 & Unit \\ 
\hline
Layer-1 & x$_{\rm low,1}$ & 0.338$^{+0.014}_{-0.014}$ & 0.990$^{u}_{-0.001}$ & 0.990$^{u}_{-0.001}$ & 0.990$^{u}_{-0.001}$ & 0.900$^{+0.003}_{l}$ & \\
& x$_{\rm high,1}$ & 0.958$^{+0.002}_{-0.002}$ & 1.000$^{u}_{-0.020}$ & 1.000$^{u}_{-0.013}$ & 1.000$^{u}_{-0.019}$ & 1.000$^{u}_{-0.008}$ & \\
& $f_{\rm nH,1}$ & 100.0$^{u}$ & 8.6$^{+1.5}_{-1.5}$ & $9.4^{+1.0}_{-1.0}$ & $9.5^{+1.0}_{-1.0}$ & $9.0^{+0.8}_{-0.8}$ & \% \\
\hline
Layer-2 & x$_{\rm low,2}$ & -- & 0.383$^{+0.012}_{-0.012}$ & 0.901$^{+0.045}_{-0.045}$ & 0.917$^{+0.027}_{-0.027}$ & 0.7-fixed & \\
& x$_{\rm high,2}$ & -- & 0.950$^{u}_{-0.001}$ & 0.929$^{+0.031}_{-0.031}$ & 0.917$^{+0.027}_{-0.027}$ & 0.9-fixed & \\
& $f_{\rm nH,2}$ & -- & 91.4$^{+2.3}_{-2.3}$ & 16.0$^{+1.7}_{-1.7}$ & 15.6$^{+1.9}_{-1.9}$ & 39.6$^{+1.5}_{-1.5}$ & \% \\
\hline
Layer-3 & x$_{\rm low,3}$ & -- & -- & 0.319$^{+0.034}_{-0.034}$ & 0.764$^{+0.118}_{-0.118}$ &0.5-fixed & \\
& x$_{\rm high,3}$ & -- & --& 0.850$^{+0.026}_{-0.026}$ & 0.765$^{+0.117}_{-0.117}$ & 0.7-fixed & \\
& $f_{\rm nH,3}$ & -- & --& 74.6$^{+7.9}_{-7.9}$ & 30.4$^{+5.3}_{-5.3}$ & 8.8$^{+5.4}_{-5.4}$ & \% \\
\hline
Layer-4 & x$_{\rm low,4}$ & -- & --& -- & 0.453$^{+0.238}_{-0.238}$ & 0.3-fixed & \\
& x$_{\rm high,4}$ & -- & --& -- & 0.453$^{+0.238}_{-0.238}$ & 0.5-fixed & \\
& $f_{\rm nH,4}$ & -- & --& -- & 44.5$^{+10.2}_{-10.2}$ & 42.6$^{+5.1}_{-5.1}$ & \% \\
\hline
Layer-5 & x$_{\rm low,5}$ & -- & -- & -- & -- & 0.1-fixed & \\
& x$_{\rm high,5}$ & -- & -- & -- & -- & 0.3-fixed & \\
& $f_{\rm nH,5}$ & -- & -- & -- & -- & 0.0$^{+2.9}_{l}$ & \% \\
\hline
Layer-6 & x$_{\rm low,6}$ & -- & -- & -- & -- & 0.0-fixed & \\
& x$_{\rm high,6}$ & -- & -- & -- & -- & 0.1-fixed & \\
& $f_{\rm nH,6}$ & -- & -- & -- & -- & 0.0$^{+1.8}_{l}$ & \% \\
\hline
& $N_{\rm H,tot}$ & 1.35$^{+0.25}_{-0.25}$ & 1.73$^{+0.05}_{-0.05}$ & 1.92$^{+0.16}_{-0.16}$ & 1.94$^{+0.23}_{-0.23}$ & 1.57$^{+0.13}_{-0.13}$ & $10^{23}$ cm$^{-2}$ \\
\hline
& $\chi^2_{\rm \nu}$ & 261.8/96 & 175.0/93 & 124.5/90 & 122.1/87 & 254.4/91 & \\
\hline
\end{tabular}
\end{minipage}
\end{table*}

\begin{table*}
 \centering
  \begin{minipage}{175mm}
  \centering
\caption{Best-fit halo models in different scenarios using the COMP-NC-B dust grain population.}
   \label{app-tab-halofit2}
\begin{tabular}{lcccccccccc}
\hline
\hline
Layer & Parameter & Scenario-1 & Scenario-2 & Scenario-3 & Scenario-4 & Scenario-5 & Unit \\ 
\hline
Layer-1 & x$_{\rm low,1}$ & 0.000$^{+0.001}_{l}$ & 0.981$^{+0.003}_{-0.003}$ & 0.988$^{+0.002}_{-0.002}$ & 0.990$^{u}_{-0.026}$ & 0.988$^{+0.002}_{-0.002}$ & \\
& x$_{\rm high,1}$ & 0.840$^{+0.006}_{-0.006}$ & 0.999$^{+0.001}_{-0.003}$ & 1.000$^{u}_{-0.020}$ & 0.995$^{+0.002}_{-0.002}$ & 1.000$^{u}_{-0.009}$ & \\
& $f_{\rm nH,1}$ & 100.0$^{u}$ & 12.0$^{+2.5}_{-2.5}$ & $14.0^{+2.1}_{-2.1}$ & $11.5^{+1.8}_{-1.8}$ & $14.2^{+2.1}_{-2.1}$ & \% \\
\hline
Layer-2 & x$_{\rm low,2}$ & -- & 0.000$^{+0.001}_{l}$ & 0.470$^{+0.032}_{-0.032}$ & 0.851$^{+0.097}_{-0.097}$ & 0.7-fixed & \\
& x$_{\rm high,2}$ & -- & 0.864$^{+0.006}_{-0.006}$ & 0.905$^{+0.008}_{-0.008}$ & 0.851$^{+0.097}_{-0.097}$ & 0.9-fixed & \\
& $f_{\rm nH,2}$ & -- & 88.0$^{+2.8}_{-2.8}$ & 32.9$^{+1.9}_{-1.9}$ & 10.8$^{+1.7}_{-1.7}$ & 16.4$^{+0.6}_{-0.6}$ & \% \\
\hline
Layer-3 & x$_{\rm low,3}$ & -- & -- & 0.000$^{+0.025}_{l}$ & 0.627$^{+0.133}_{-0.133}$ &0.5-fixed & \\
& x$_{\rm high,3}$ & -- & --& 0.000$^{+0.025}_{l}$ & 0.627$^{+0.133}_{-0.133}$ & 0.7-fixed & \\
& $f_{\rm nH,3}$ & -- & --& 53.1$^{+4.1}_{-4.1}$ & 22.3$^{+4.7}_{-4.7}$ & 14.1$^{+1.4}_{-1.4}$ & \% \\
\hline
Layer-4 & x$_{\rm low,4}$ & -- & --& -- & 0.000$^{+0.022}_{-0.022}$ & 0.3-fixed & \\
& x$_{\rm high,4}$ & -- & --& -- & 0.000$^{+0.022}_{-0.022}$ & 0.5-fixed & \\
& $f_{\rm nH,4}$ & -- & --& -- & 55.4$^{+10.0}_{-10.0}$ & 0.0$^{+2.9}_{l}$ & \% \\
\hline
Layer-5 & x$_{\rm low,5}$ & -- & -- & -- & -- & 0.1-fixed & \\
& x$_{\rm high,5}$ & -- & -- & -- & -- & 0.3-fixed & \\
& $f_{\rm nH,5}$ & -- & -- & -- & -- & 0.0$^{+3.1}_{l}$ & \% \\
\hline
Layer-6 & x$_{\rm low,6}$ & -- & -- & -- & -- & 0.0-fixed & \\
& x$_{\rm high,6}$ & -- & -- & -- & -- & 0.1-fixed & \\
& $f_{\rm nH,6}$ & -- & -- & -- & -- & 55.3$^{+1.9}_{-1.9}$ & \% \\
\hline
& $N_{\rm H,tot}$ & 1.47$^{+0.02}_{-0.02}$ & 2.15$^{+0.08}_{-0.08}$ & 2.61$^{+0.13}_{-0.13}$ & 2.56$^{+0.29}_{-0.29}$ & 2.58$^{+0.14}_{-0.14}$ & $10^{23}$ cm$^{-2}$ \\
\hline
& $\chi^2_{\rm \nu}$ & 610.0/96 & 504.4/93 & 114.1/90 & 111.1/87 & 119.6/91 & \\
\hline
\end{tabular}
\end{minipage}
\end{table*}

\end{document}